\documentstyle[epsfig,referee]{mn}

\newif\ifAMStwofonts
\newcommand{\bvec}[1] {\mbox{\boldmath$ #1$}}
\newcommand{\be} {\begin{equation}}
\newcommand{\ee} {\end{equation}}
\def\witchbox#1#2#3{\hbox{$\mathchar"#1#2#3$}}
\def\leqsim{\mathrel{\rlap{\lower3pt\witchbox218}\raise2pt\witchbox13C}}
\def\geqsim{\mathrel{\rlap{\lower3pt\witchbox218}\raise2pt\witchbox13E}}

\ifoldfss
  \ifCUPmtlplainloaded \else
    \NewTextAlphabet{textbfit} {cmbxti10} {}
    \NewTextAlphabet{textbfss} {cmssbx10} {}
    \NewMathAlphabet{mathbfit} {cmbxti10} {} % for math mode
    \NewMathAlphabet{mathbfss} {cmssbx10} {} %  "   "    "
  \fi
  \ifAMStwofonts
    \ifCUPmtlplainloaded \else
      \NewSymbolFont{upmath} {eurm10}
      \NewSymbolFont{AMSa} {msam10}
      \NewMathSymbol{\upi}     {0}{upmath}{19}
      \NewMathSymbol{\umu}     {0}{upmath}{16}
      \NewMathSymbol{\upartial}{0}{upmath}{40}
      \NewMathSymbol{\leqslant}{3}{AMSa}{36}
      \NewMathSymbol{\geqslant}{3}{AMSa}{3E}

       \let\le=\leqslant
       
    \fi
  \fi
\fi % End of OFSS

\ifnfssone
  \newmathalphabet{\mathit}
  \addtoversion{normal}{\mathit}{cmr}{m}{it}
  \addtoversion{bold}{\mathit}{cmr}{bx}{it}
  \newmathalphabet{\mathbfit} % math mode version of \textbfit{..}
  \addtoversion{normal}{\mathbfit}{cmr}{bx}{it}
  \addtoversion{bold}{\mathbfit}{cmr}{bx}{it}
  \newmathalphabet{\mathbfss} % math mode version of \textbfss{..}
  \addtoversion{normal}{\mathbfss}{cmss}{bx}{n}
  \addtoversion{bold}{\mathbfss}{cmss}{bx}{n}
  \ifAMStwofonts
    \ifCUPmtlplainloaded \else
      \UseAMStwoboldmath
      \makeatletter
      \new@mathgroup\upmath@group
      \define@mathgroup\mv@normal\upmath@group{eur}{m}{n}
      \define@mathgroup\mv@bold\upmath@group{eur}{b}{n}
      \edef\UPM{\hexnumber\upmath@group}
      \new@mathgroup\amsa@group
      \define@mathgroup\mv@normal\amsa@group{msa}{m}{n}
      \define@mathgroup\mv@bold\amsa@group{msa}{m}{n}
      \edef\AMSa{\hexnumber\amsa@group}
      \makeatother
      \mathchardef\upi="0\UPM19
      \mathchardef\umu="0\UPM16
      \mathchardef\upartial="0\UPM40
      \mathchardef\leqslant="3\AMSa36
      \mathchardef\geqslant="3\AMSa3E

       \let\le=\leqslant

    \fi
  \fi
\fi % End of NFSS release 1

\ifnfsstwo
  \DeclareMathAlphabet{\mathbfit}{OT1}{cmr}{bx}{it}
  \SetMathAlphabet\mathbfit{bold}{OT1}{cmr}{bx}{it}
  \DeclareMathAlphabet{\mathbfss}{OT1}{cmss}{bx}{n}
  \SetMathAlphabet\mathbfss{bold}{OT1}{cmss}{bx}{n}
  \ifAMStwofonts
    \ifCUPmtlplainloaded \else
      \DeclareSymbolFont{UPM}{U}{eur}{m}{n}
      \SetSymbolFont{UPM}{bold}{U}{eur}{b}{n}
      \DeclareSymbolFont{AMSa}{U}{msa}{m}{n}
      \DeclareMathSymbol{\upi}{0}{UPM}{"19}
      \DeclareMathSymbol{\umu}{0}{UPM}{"16}
      \DeclareMathSymbol{\upartial}{0}{UPM}{"40}
      \DeclareMathSymbol{\leqslant}{3}{AMSa}{"36}
      \DeclareMathSymbol{\geqslant}{3}{AMSa}{"3E}

       \let\le=\leqslant

    \fi
  \fi
\fi % End of NFSS release 2

\ifCUPmtlplainloaded \else
  \ifAMStwofonts \else % If no AMS fonts
    \def\upi{\pi}
    \def\umu{\mu}
    \def\upartial{\partial}
  \fi
\fi

\title{On the  tidal interaction
of massive  extra-solar planets on highly eccentric orbit }
\author[P. B. Ivanov, J. C. B. Papaloizou]
       {P. B. Ivanov 
\\ Astronomy Unit, School of Mathematical Sciences, Queen Mary,
University of London, UK \\
Astro Space Center of PN Lebedev Physical 
Institute, Moscow, Russia
\newauthor J. C. B. Papaloizou
\\ Astronomy Unit, School of Mathematical Sciences, Queen Mary,
University of London, UK}

\pagerange{\pageref{firstpage}--\pageref{lastpage}}
\pubyear{1994}

\begin{document}

\maketitle

\label{firstpage}

\begin{abstract}
In this paper we develop a theory of disturbances induced  by the stellar tidal field in a  fully convective 
slowly rotating planet orbiting on a highly eccentric orbit around a central star.  
In that case it is appropriate to treat the tidal influence as a succession 
of  impulsive tidal interactions occurring at  periastron passage.   For a fully convective planet 
mainly  the $l=2$ fundamental mode of oscillation is excited. We show that  there are 
two contributions
to the mode energy and angular momentum gain due to impulsive  tidal interaction:
a) 'the quasi-static' contribution which requires
dissipative processes operating in the planet; b) the  dynamical  contribution associated with 
excitation of modes of oscillation. These   contributions are obtained self-consistently 
from a single set of the governing equations.
We calculate a critical 'equilibrium' value of angular velocity of the planet 
$\Omega_{crit}$ determined by the condition that  action of the dynamical tides does not alter  
the angular velocity at that rotation rate.  We show that this can be much larger
than the  corresponding  rate  associated with  quasi-static tides and that at this  angular velocity,
the rate of energy exchange is minimised.
We also investigate the conditions for the stochastic increase in oscillation energy
that may occur if many periastron passages are considered and dissipation is not important.
We provide a simple criterion
for this  instability to occur.

Finally, we make some simple estimates of time scale of evolution of the orbital semi-major
axis and circularization of initially eccentric orbit 
due to tides, using a realistic model of the planet and its cooling history,
for orbits with  periods after circularization typical of those observed for  extra-solar planets 
$P_{obs} \geqsim 3days$. 
Quasi-static tides  are found to be ineffective for semi-major axes $\geqsim 0.1Au$. 
On the other hand,  dynamic tides  could  have produced
a very large  decrease of the semi-major axis of a planet with mass of the order of the Jupiter mass $M_{J}$
and   final period $P_{obs}\sim 1-4.5days$  on a time-scale $\leqsim$ a few Gyrs. In that case the original
semi-major axis may be as large as $\geqsim 10^{2}Au.$

For  planets with masses $\geqsim 5M_{J}$  dynamical tides  excited in the
star appear to be more important than the tides excited  in the planet. 
They may also, in principle,  result in orbital evolution in a time less than or comparable to the life time
of the planetary systems.
Finally we point out that there are several issues in the context of
the scenario of the circularization of the orbit solely due to dynamic tides that remain to be resolved.
Their possible resolution is discussed.

\end{abstract}

\begin{keywords}
tides, binary star, stellar rotation, planetary systems: formation

\end{keywords}

\section{Introduction}

The recently  discovered  100 or so extra-solar giant planets
have masses  in the range
$0.12 - 11 $ Jupiter masses. They   may be found
at distances of several $Au$ or
close to the central star with periods of a few days. High orbital
eccentricities are common
(e.g. Mayor \& Queloz 1995;
 Marcy \& Butler 1998; Marcy et al 2000).

It has been suggested  that  giant planets may form through gravitational
instability in a disc at large radii (e.g.. Boss 2002) or through
the  `critical core mass' model in which
a critical
$\sim 15$ M$_{\oplus}$ core is formed in a disc  by
accumulation of solids and then undergoes rapid
gas
accretion (see e.g. Papaloizou, Terquem \& Nelson 1999,  for a review
and appropriate references).
In this case it is expected that the cores of
gas giant planets should begin to form beyond a radius of $r \sim 4$ $Au$, the
so--called `ice condensation radius' where the existence
of ices facilitates the accumulation of solids.

In order to explain the existence of
the closely orbiting extra-solar giant planets
one is then led to propose
orbital migration.  In principle this may occur
through the gravitational interaction between
a protoplanet and the protostellar disc (e.g. Lin \& Papaloizou 1986, Nelson et al 2000 
and references therein)
or through mutual
gravitational interactions among a strongly interacting system
of protoplanets, and further circularization by tides. In this Paper we explore
the second possibility.

The presence of high orbital eccentricities amongst extra solar
planets is suggestive of a strong orbital relaxation
or scattering process.
The gaseous environment of a disc
may act to inhibit such interactions until it is removed.
Gas free dynamical interactions of  coplanar protoplanets formed on
 neighbouring circular orbits have been considered by
Weidenschilling \& Mazari (1996) and  Rasio \& Ford (1996).
These may produce close scattering and high eccentricities
but the observed distribution of extra solar planets
is not reproduced.

Papaloizou \& Terquem (2001) investigate  a scenario in which
$5 \le N \le 100$ planetary objects  in the  range
of several Jupiter masses
are assumed
to form  rapidly through  fragmentation  or gravitational
instability occurring
a disc  or protostellar envelope on a scale of $R_{max}= 100Au$.
If these objects are put down in circular orbits
about a solar mass star, at random
in a volume contained
within a spherical shell with inner and outer radii of $0.1R_{max}$ and
$R_{max}$ respectively, the system relax
on a time-scale $\sim 100$ orbits and the final state of the system is
independent of details of initial conditions.
In fact, the evolution is similar to that of a stellar cluster, most objects
escape leaving at most 3  bound planets.
Close encounters or collisions  with the central star occurred
for about 10\% of cases in which a planet  goes through a phase where
it is in a highly eccentric orbit that has close approaches
to the central star.

These simulations naturally lead us to a situation
where
tidal interaction leading to orbital
circularization
and  to the formation of a very closely orbiting giant planet
may become a possibility. In this Paper we analyse the tidal interactions in detail.
For simplicity we neglect the influence of gravitational interactions between the planets
and consider only the tidal interactions between the planet and the central star.

We develop the theory of disturbances induced by 
tides in a giant planet in Section 2.
We consider the case of a  highly eccentric orbit and a 
fully convective slowly rotating planet
taking into account all the 
specific features of this situation.  This is such that the theory of tidal
disturbances is sufficiently simple that
it is possible to obtain compact analytical expressions for the energy and  angular 
momentum gained by  the planet after  a close encounter
with the star, see equations (44,45,47) for quasi-static tides associated with frictional 
processes in the planet  and (60-65) for dynamic tides associated with excitation of the
oscillation modes.

It has been mentioned by a number of authors (Press $\&$ Teukolsky 1977, hereafter PT, Kochanek 1992,
Mardling 1995 a,b) that when there are  many periastron passages,
energy can  flow backwards and forwards between the
orbit  and oscillation mode. This happens 
in a situation where the  dissipation time of
the normal mode is larger than the orbital period of the planet.
The direction of energy transfer resulting from a particular periastron passage
depends on the phase of the  pulsation at the
moment of  periastron passage.
We address this multiple passage problem in Section 2.9  and show that
when the semi-major axis of the orbit is
 sufficiently large, the energy transfer is highly irregular as  has
been  previously  found numerically
(Mardling 1995 a,b).
Here we find a simple semi-analytic criterion for  such
irregular behaviour. When this criterion is fulfilled the mode energy grows on average proportional to
the number of periastron passages and we can use
calculations based on 'naive' adding of the energy gained from the
orbit by the normal mode in order to make estimates of the evolution time-scale.
The similar situation occurs if one simply assumes
that the energy transferred to normal modes is dissipated between
successive periastron passages.

In Section 3 we adopt the above approach 
to calculate time-scales required for significant 
orbital circularization due to tides.
Finally in Section 4 we discuss our results and present conclusions.

\section{ The tidal interaction of a rotating convective planet
in a weakly bound orbit}

In this Section we consider the tidal perturbations induced
in a rotating planet after a  parabolic encounter
with a central star. We derive expressions for the 
energy and angular momentum gained
by the planet  as a result of  such an encounter.
By considering 
successive encounters,
the results can be used to discuss planets on highly eccentric orbits
as well as parabolic ones.
In the absence of explicit 
dissipation  only the excitation of normal modes
or dynamic tides play a role. 
For a non-rotating planet
the energy gain has been calculated by 
PT. The equations for the 
perturbations  induced in a rotating object have also been discussed
(e.g. Lai 1997 and references therein). However,
only dynamic tides were discussed and  the treatment of viscous terms in
the equations of motion was incorrect.
Stationary  or quasi-static tides for very eccentric orbits 
have been considered by Alexander (1973), Hut (1981), Zahn (e.g. Zahn 1989 and references therein) 
and more recently by Eggleton, Kiseleva and Hut (1998)
using a very different approach based on calculating the
phase lag of an assumed quasi-static tidal bulge. Here we
demonstrate how the expressions for 
energy and angular momentum gain  associated with the
the  quasi static and resonant  dynamic 
tides follow from   a single formulation of the problem
leading to one governing equation.
 
In general, the linearised equations  governing  small perturbations of a
rotating object are non-separable.   However, for a fully convective planet and 
sufficiently slow rotation the angular velocity $\Omega_r$
is much smaller than the internal
'dynamical' frequency $\Omega_{*}=\sqrt{{Gm_{pl}\over R_{pl}^{3}}}$, where
$m_{pl}$ is the mass of the planet. Then it is possible to construct a
self-consistent theory of perturbations  based on 
an expansion in the  small parameter $\Omega_{r}/\Omega_{*}$
\footnote{Note that this condition is approximately valid even for a fast rotator like Jupiter. The ratio of Jupiter's
rotation frequency to $\Omega_{*}$ is about $0.28$  which 
can be  considered as small parameter in 
perturbation  theory 
(see e.g. Papaloizou $\&$ Pringle 1978, Lee 1993
for more discussion of this point).}. 
Here, only zeroth and first order terms in
that expansion will
be taken into account. 

We introduce the inner product (e.g. PT) 
\be <\bvec {\xi}_1 | \bvec {\xi}_2>=\int d^3x \rho (\bvec{\xi}_1\cdot \bvec{\xi}_2), \label{eqno1}\ee   
where $(\bvec {\xi}_1\cdot \bvec { \xi}_2)$ is the scalar product of two vectors,
and $\rho$ is the density. For simplicity 
we consider only the  $l=2$  component of the  tidal perturbation
which dominates for large orbital separations. Further  
since $g$ modes are absent
in the fully  convective objects we consider
and the frequencies of the 
$p$ modes are always significantly
larger than the frequency of the $f$ mode making
their tidal excitation relatively small, we take account only of
the fundamental mode when calculating the tidal response.
\footnote{An extension of the analysis 
to take $p$ modes into account is straightforward.}.

\subsection{Basic equations}
We use the  linearised equations of motion  governing small perturbations 
in the  inertial frame.
These have been discussed and  used by many authors (e.g.. Lynden-Bell \& Ostriker 1967,
Papaloizou \& Pringle 1978) so we do not present a derivation here
writing them in the form:

\be {\partial^{2}\over \partial t^{2}}\bvec {\xi}+
2(\bvec { v}_0\cdot \nabla ){\partial \over \partial t}\bvec { \xi} 
+{\bf C}(\bvec {\bf \xi})=\nabla U+ {\bf f}_{\nu}. \label{eqno2}\ee   
Here $\bvec {\xi}$ is the Lagrangian displacement, ${\bf C}(\bvec {\xi})$ is the linear 
self-adjoint operator accounting for the action
of pressure and self-gravity on perturbations (Chandrasekhar, 1964, Lynden-Bell $\&$ Ostriker 1967). $U$ is the tidal
potential given by
\be U=GM\left({1\over |{\bf r}-\bvec {\bf D}(t)|} -{ {\bf r}\cdot{\bf D}(t) \over 
 D(t)^3} \right)
\approx {GM\over D(t)}\left[ \sum_{m=-2,0,2} W_m {\left({r \over D(t)}\right )}^{2} 
e^{-im\Phi(t)}Y_{2m}(\theta, \phi)\right ] . \label{eqno3}\ee
Here $(r, \theta, \phi)$ are  spherical coordinates with origin at the center of the planet
and  polar axis coinciding with the rotational axis, all angular momenta being assumed aligned.
The mass of the star is $M,$ ${\bf D}(t)$ is  the position vector of
 the star assumed to  orbit in the plane $\theta = \pi/2$   with 
$D(t)=|{\bf D}(t)|$  and $\Phi(t)$  being  the associated azimuthal angle 
at some arbitrary moment of time
$t.$ The moment $t=0$ is chosen to correspond to closest approach of star and planet. For $|m|=2,$
$W_m= \sqrt{3\pi/10}$  and 
$W_0 =-\sqrt{\pi/5},$ with $Y_{lm}(\theta, \phi)$  being the usual  spherical harmonic.
Note, that in 
equation (3) the indirect term due to the acceleration of the coordinate
system is included  as the second term on the left hand side.   
On the right hand side we take into account only the leading term in the
expansion in powers of $r/D(t)$. 

The unperturbed steady state  velocity field is 
$ {\bf v}_0.$   For the problem considered here,
 this is associated with uniform  rotation of the planet such that
\be {\bf v}_0=\bvec { \Omega}_{r}\times  {\bf r}. \label{eqno4}\ee 
The viscous force per unit  mass is
$ {\bf f}_{\nu}.$  This may be written in component form in Cartesian coordinates  as 
\be f^{\alpha}_{\nu}=(1/\rho)t^{\alpha \beta}_{ ,\beta}=(1/\rho)(\rho \nu (v^{\alpha}_{ ,\beta}+
v^{\beta}_{ ,\alpha}-(2/3)v^{\gamma}_{ ,\gamma}\delta^{\beta}_{\alpha}))_{ ,\beta}, \label{eqno5} \ee
where the comma stands for differentiation and the usual
summation convention over the Greek indices is adopted from now on. 

The components of
the viscous tensor are 
$t^{\alpha \beta}$ 
and $\nu$ is the kinematic viscosity. Since we assume  uniform rotation of the
unperturbed planet, the unperturbed  velocity field (\ref{eqno4}) does not contribute 
to equation (\ref{eqno5})  so that
the velocity components $v^{\alpha}$ may be considered to be  solely due to the
perturbations and  are thus related to $\bvec { \xi}$  through 
\be {\bf v}={\partial \over \partial t}\bvec {\xi}+\Omega_{r}{\partial \over \partial \phi}\bvec { \xi}. \label{eqno6}\ee

\subsection{Normal Modes of the Rotating Planet}

The solution for the perturbations of a planet occurring
as a result of an encounter with a central star  will naturally involve its
normal modes $\bvec{\xi}_k$ with their associated eigenfrequencies
$\omega_{k}.$  These satisfy
\be -\omega_{k}^{2}\bvec {\xi}_{k}
-2i\omega_{k}( {\bf v}_0\cdot \nabla )\bvec{\xi}_{k}+{\bf C}(\bvec {\xi}_{k})=0, \label{eqno7}\ee
with the standard boundary conditions (e.g. Lynden-Bell \& Ostriker 1967).

When the flow ${\bf v}_0$ is absent the eigenfunctions $\bvec {\xi}_{k}$
 reduce to $\bvec {\xi}_{k0}$   and they become those appropriate
to a spherical star (e.g. Tassoul 1978). The
associated eigenvalues  are then $\omega_{k0}.$ 
The $\bvec {\xi}_{k0}$ may then  be assumed to be orthogonal in terms of the product (\ref{eqno1}) and
normalised by the standard condition
\be <\bvec { \xi}^{*}_{k0} |\bvec { \xi}_{k0}>=1, \label{eqno8} \ee 
In this non rotating limit the eigenfunctions have  a simple dependence
on $\phi$ through a factor $\exp(im\phi)$ with $m$ being the azimuthal mode number.
We do not show this dependence explicitly here.

As we use an expansion in the  small parameter $\Omega_{r}/\Omega_{*}$:
$\bvec {\xi}_{k}=\bvec { \xi}_{k0}+\bvec {\xi}_{k1}$ and
$\omega_{k}=\omega_{k0}+\omega_{k1}$, where $\bvec {\xi}_{k1}$ and the frequency
correction  $\omega_{k1}$ are 
first order in the small parameter.
The frequency correction has the standard form (e.g. Tassoul 1978;  Christensen-Dalsgaard 1998 and references therein)
\be \omega_{k1}=-i<{\bf \xi}^{*}_{k0}|({\bf v}_0\cdot \nabla ){\bf \xi}_{k0}>. \label{eqno9}\ee

\subsection{Eigenfunction Expansion}
We look for a solution to equation (\ref{eqno2}) in the form
\footnote{Adding the corrections $\bvec { \xi}_{k1}( {\bf r})$ to $\bvec { \xi}_{k0}( {\bf r})$ in equation (10)
does not alter our results.}
\be \bvec { \xi}=\sum_{k} b(t)_{k}\bvec { \xi}_{k0}( {\bf r}). \label{eqno10}\ee

\noindent The eigenfunctions satisfy the  standard  equation for a non-rotating object:
\be -\omega_{k0}^{2}\bvec {\xi}_{k0}+{\bf C}(\bvec { \xi}_{k0})=0. \label{eqno11}\ee
We shall ultimately retain  only those corresponding  to the $l=2$ fundamental mode and
there are three of these corresponding to azimuthal mode number $m= -2,0,2.$
We use the standard representation
\be \bvec {\xi}_{k0}=\xi(r)_{R}Y_{2m}(\theta, \phi) {\bf e}_{r}+\xi(r)_{S}(r\nabla Y_{2m}(\theta, \phi)). \label{eqno11a}\ee
The components $\xi(r)_{R}$ and $\xi(r)_{S}$  are real and independent
of $m,$  but  not independent of each other.
For a convective isentropic non-rotating planet the motion 
associated with a mode  must be circulation  free.
Accordingly $\nabla \times \bvec {\xi}_{k0} =0,$ which leads to
\be {d\over dr}\xi_{S}={\xi_{R}-\xi_{S}\over r}. \label{eqno12}\ee

\subsection{Solution of the Tidal Problem}
Substituting equation (\ref{eqno10}) into equation (\ref{eqno2}) and using equations 
(\ref{eqno3} - \ref{eqno6}) and equations (\ref{eqno8} - \ref{eqno11})
with the orthogonality of the eigenfunctions,
we obtain  under the assumptions of small unperturbed rotation rate, small viscous forces
and dominance of the $f$ modes:
\be \ddot b_{k}+\omega^{2}_{k0}b_{k}+2i\omega_{k1}\dot b_{k}+\gamma(\dot b_{k}+im\Omega_{r}b_{k})=f^{T}_{k}. \label{eqno13}\ee
There are three equations of the above form for the modes with $m= -2, 0, 2$
respectively. From now on we  accordingly find it convenient to 
make the equivalence $ k \equiv m.$

The viscous damping rate $\gamma$  has the form
\be \gamma={1\over 2}\int d^3x \nu \rho \sigma^{*\alpha}_{\beta}\sigma^{\beta}_{\alpha}, \label{eqno14}\ee
where
\be \sigma^{\alpha}_{\beta}=\xi^{\alpha}_{ ,\beta}+\xi^{\beta}_{ ,\alpha}-(2/3)\xi^{\gamma}_{ ,\gamma}\delta^{\alpha}_{\beta}. \label{vstr} \ee  
One can readily verify that for a normalised eigenfunction, $\gamma^{-1}$
has the dimension of  time and corresponds to a viscous diffusion
time across the planet.

The quantity $f_{k}^{T}$ determines the tidal coupling  and is given by
\be  f_{k}^{T}=<\bvec { \xi}^{*}_{k0}|\nabla U>. \label{eqno15} \ee
The coefficients in equation (\ref{eqno13})
can be obtained in explicit form.
Substituting equations (4), (11) into equation (14)  we obtain
(e.g. Christensen-Dalsgaard 1998 and references therein) 
\be \omega_{k1}=m\beta \Omega_{r}, \label{eqno16}\ee 
where
\be \beta=1-\int^{R_{pl}}_{0}r^{2}dr\rho(2\xi_{R}\xi_{S}+\xi_{S}^2). \label{eqno16ab}\ee
An explicit expression for the damping rate $\gamma $ is derived in Appendix. The result is
\be \gamma=4\int^{R_{pl}}_{0} r^{2}dr\rho \nu \lbrace {1\over 3} (\xi_{R}^{'}-{\xi_{R}\over r}+
3{\xi_{S}\over r})^{2}+
{1\over r^{2}}({(\xi_{R}-3\xi_{S})}^{2}+2\xi_{R}^{2}+3{(\xi_{R}-\xi_{S})}^{2})\rbrace, 
\label{eqno17}\ee 
where the prime stands for differentiation with respect to $r$.
Finally, the quantity $f_{k}^{T}$ has the form
\be  f_{k}^{T}={GMQ\over D(t)^{3}}W_{m}e^{-im\Phi(t)}, \label{eqno18}\ee
where the overlap integral $Q$ is expressed as (PT)
\be Q=2\int^{R_{pl}}_{0}dr\rho r^{3}(\xi_{R}+3\xi_{S}). \label{eqno19} \ee 

Equation (\ref{eqno13})
 is close to that  obtained by Lai (1997). The only difference  relates to
the treatment of the the last  term on
the left hand side. We give the explicit 
expression for the damping rate $\gamma$ (see  equation (\ref{eqno17})).  
Note the contribution proportional to  $(im\Omega_{r}b_{m})$ in equation (\ref{eqno13}).
This  arises from   converting     
Eulerian velocity perturbations to Lagrangian displacements.  It is necessary to
include this term  in order to obtain 
correct expressions for the energy and angular momentum
exchange between planet and orbit associated with quasi static tides (see below).

It is convenient to introduce the real quantities 
\be a_{+}={1\over \sqrt 2}(b_{2}+b_{-2}),\quad a_{-}={i\over \sqrt 2}(b_{2}-b_{-2}),\quad a_{0}=b_{0}, \label{eqno 21}\ee
and 
\be f^{T}_{+}={1\over \sqrt 2}(f_{2}^{T}+f_{-2}^{T})=\sqrt{2}{GMQ\over D(t)^{3}}W_{2}\cos{(2\Phi(t))}, \label{eqno22}\ee
\be f^{T}_{-}={i\over \sqrt 2}(f_{2}^{T}-f_{-2}^{T})
=\sqrt{2}{GMQ\over D(t)^{3}}W_{2}\sin{(2\Phi(t))}, \label{eqno23}\ee 
and rewrite  equation (\ref{eqno13}) in terms of them to obtain
\be \ddot a_{+}+\omega^{2}_{00}a_{+}+ \sigma \dot a_{-}
+\gamma(\dot a_{+}+ 2\Omega_{r}a_{-})=f^{T}_{+}, \label{eqno24}\ee   
\be \ddot a_{-}+\omega^{2}_{00}a_{-}- \sigma \dot a_{+}
+\gamma(\dot a_{-}- 2\Omega_{r}a_{+})=f^{T}_{-}, \label{eqno25}\ee
\be \ddot a_{0}+\omega^{2}_{00}a_{0}+\gamma \dot a_{0}=f^{T}_{0}, \label{eqno26}\ee  
where
$\sigma=2|\omega_{k1}|$.

\subsection{Energy and Angular Momentum Exchanges}
As we wish to consider tidally induced exchanges of energy
and angular momentum between planet and orbit, we  first consider
the energy and angular momentum associated with the perturbations.

\noindent The canonical energy $E$ and angular momentum $L$ 
of perturbations can be expressed as (e.g. Friedman and Schutz 1978)
\be E={1\over 2}({\dot a_{0}}^{2}+{\dot a_{+}}^{2}+{\dot a_{-}}^{2})+
{\omega_{00}^2\over 2}(a_{0}^{2}+a_{+}^{2}+a_{-}^{2}), \label{eqno26a}\ee
and
\be L=2(a_{+}\dot a_{-}-a_{-}\dot a_{+}-{\sigma\over 2}(a_{+}^{2}+a_{-}^{2})). \label{eqno26b}\ee
Equations governing their time evolution are readily found from equations (\ref{eqno24} 
- \ref{eqno26}).
Writing ${dE \over dt} = {d E^{T}\over dt} + {d E^{\nu}\over dt}$, 
and ${d L\over dt} = {d L^{T}\over dt} + {d L^{\nu}\over dt}$,  the result can be expressed
in the form

\be {d E^{T} \over dt}=\dot a_{0}f^{T}_{0}+\dot a_{+}f^{T}_{+}+\dot a_{-}f^{T}_{-}, \label{eqno27}\ee
\be {d E^{\nu} \over dt}=-\gamma ({\dot a_{0}}^{2}+{\dot a_{+}}^{2}+{\dot a_{-}}^{2}-
2\Omega_{r}(a_{+}\dot a_{-}-a_{-}\dot a_{+})), \label{eqno28}\ee
\be {d L^{T} \over dt}=2(a_{+}f^{T}_{-}-a_{-}f^{T}_{+})  \label{eqno29}\ee  and 
\be {d L^{\nu}\over dt}=-2\gamma (a_{+}\dot a_{-}-a_{-}\dot a_{+}-2\Omega_{r}(a_{+}^{2}+a_{-}^{2})). \label{eqno30}\ee 

One may regard 
${d E^{T} / dt}$, ${d L^{T}/ dt},$  ${d E^{\nu}/ dt}$, ${d L^{\nu} / dt}$
as determining the  increase  of the perturbation energy and angular momentum
arising from tides and viscosity . These quantities
can be easily calculated with the help 
of equations (\ref{eqno24} - \ref{eqno26}).

\subsection{Quasi-Static Tides}
In the limit when the time-scale associated with changes at closest
approach is much longer  than  the internal dynamical time-scale,  the planet
responds by evolving through a sequence of quasi-hydrostatic equilibria.
In this approximation,  significant perturbations are only induced by tides in the vicinity
of closest approach. 
 Energy and angular momentum transfer
to the planet occurs only  by  the 
 action of viscosity. 

The energy and angular momentum gained by the planet  
is equal to that dissipated in perturbations and can be expressed as
\be \Delta E_{st}=-\int^{+\infty}_{-\infty}dt{d E^{\nu}\over dt}, \label{eqno31}\ee
\be \Delta L_{st}=-\int^{+\infty}_{-\infty}dt{d L^{\nu}\over dt}, \label{eqno32}\ee
where $t=0$ corresponds to the moment of closest approach. 
To  evaluate the above integrals, we look for the  solution to equations 
 (\ref{eqno24} - \ref{eqno26}) in the form of an expansion in terms of the small
parameter $\Omega_{p}/\Omega_{*}$, where
\be \Omega_{p}=\sqrt{GM\over D_{min}^{3}}  \label{eqno33}\ee
and $D_{min}$ is the value of periastron distance. 

We also assume that $\omega_{00}\sim \Omega_{*}$ and
$\Omega_{r}\sim \Omega_{p}$. In the leading approximation the solution is trivial:
\be a_{0}={f_{0}^{T}\over \omega_{00}^{2}},
 \quad a_{\pm}={f_{\pm}^{T}\over \omega_{00}^{2}}. \label{eqno34}\ee
Substituting (\ref{eqno34}) into (\ref{eqno28}) and (\ref{eqno30}), and evaluating the 
integrals in  (\ref{eqno31}) and (\ref{eqno32}) 
we obtain

\be \Delta E_{st}={\gamma \over \omega_{00}^{4}}(I_{1}-2\Omega_{r}I_{2}), \quad
\Delta L_{st}={2\gamma \over \omega_{00}^{4}}(I_{2}-2\Omega_{r}I_{3}), \label{eqno35}\ee
where
$$I_{1}=\int^{+\infty}_{-\infty}dt ({(\dot f^{T}_{0})}^{2}+{(\dot f^{T}_{+})}^{2}+{(\dot f^{T}_{-})}^{2}),
\quad I_{2}=\int^{+\infty}_{-\infty}dt (f_{+}^{T}\dot f^{T}_{-}-f_{-}^{T}\dot f^{T}_{+}), $$ 
\be I_{3}=\int^{+\infty}_{-\infty}dt ({(f^{T}_{+})}^{2}+{(f^{T}_{-})}^{2}). \label{eqno36}\ee

For a parabolic orbit, the evaluation of the integrals in  equations (\ref{eqno36}) is straightforward. After elementary,
but rather tedious calculations we obtain 
\be \Delta E_{st}=\beta_{1}{\gamma \Omega_{p}\over \omega_{00}^{4}}
{{(GMQ)}^{2}\over D_{min}^{6}}
(1-{112\over 117}{\Omega_{r}\over \sqrt 2 \Omega_{p}}), \label{eqno36a}\ee 
\be \Delta L_{st}=\beta_{2}{\gamma \over \omega_{00}^{4}}{{(GMQ)}^{2}\over D_{min}^{6}} 
(1-{40\over 33}{\Omega_{r}\over \sqrt 2 \Omega_{p}}), \label{eqno37}\ee
where $\beta_{1}={11583\over 5120}\sqrt 2\pi^{2}\approx 31,6$, $\beta_{2}={693\over 320}\pi^{2}\approx 21,4$. 
It is very helpful to introduce new dimensionless 
variables in equations (\ref{eqno36a} - \ref{eqno37}).
Instead of using $D_{min}$ directly,
we  specify the closest approach distance
by use of  the Press-Teukolsky   parameter $\eta$
defined as
\be \eta={\Omega_{*}\over \Omega_{p}}=
\sqrt{ {m_{pl}\over M}{D_{min}^{3}\over R_{pl}^{3}}}. \label{eqno38}\ee
The linear theory of perturbations is valid only for sufficiently large values of $\eta$.
As we will see in the next Section, in our problem we have typically $\eta \sim 10$, and the
use of the linear theory is fully justified.

We  find it convenient to 
express $Q$, $\omega_{00}$, $\gamma$, $\Omega_{r}$ in terms of  natural units. 
In so doing we introduce dimensionless variables indicated with a tilde.
These are specified  through
$Q=\tilde Q\sqrt m_{pl} R_{pl}$, $\omega_{00}=\tilde \omega_{00}\Omega_{*}$,
$\gamma=\tilde \gamma \Omega_{*}$, $\Omega_{r}=\tilde \Omega_{r} \Omega_{*}$.  
Note that $\gamma^{-1},$ being related to the internal viscous
diffusion time  is expressed in units of the internal dynamical
time. Thus although it is dimensionless, $\tilde \gamma $
may be a very small quantity. Using the dimensionless variables, the energy and angular momentum changes may be expressed as

\be \Delta E_{st}=\beta_{1}{\tilde \gamma \tilde Q^{2}\over \tilde \omega_{00}^{4}}{E_{pl}\over \eta^{5}}
(1-{112\over 117}{\eta \tilde \Omega_{r}\over \sqrt 2}), \label{eqno39}\ee
and
\be \Delta L_{st}=\beta_{2}{\tilde \gamma \tilde Q^{2}\over \tilde  \omega_{00}^{4}}{L_{pl}\over \eta^{4}} 
(1-{40\over 33}{\eta \tilde \Omega_{r}\over \sqrt 2 }). \label{eqno40}\ee
Here these are normalised in terms of the energy and angular momentum scales 
$E_{pl}=Gm_{pl}^{2}/R_{pl}$  and  $L_{pl}=m_{pl}\sqrt{Gm_{pl}R_{pl}}$ respectively.
However, these values may only be approached for very close encounters
and comparable viscous and dynamical times.

Equations  (\ref{eqno39}) and (\ref{eqno40})
 are formally identical to the equations obtained by Hut (1981)
who used a weak friction model in the limit of the planet orbit being marginally
bound to the star,
 provided that we make  the identification
\be 
{k\over \tilde T}={5\over 8\pi}{\tilde \gamma \tilde Q^{2}\over \tilde \omega_{00}^{4}},
 \label{eqno41} \ee    
where $k$ is the apsidal motion constant, $\tilde T=1/(\Omega_{*}\tau)$,  with 
$\tau$  being the time lag
between the tidal forcing and the response of the planet that occurs through
frictional processes.

The  advantage of our approach 
is that it allows us to obtain the correct expressions for the  energy and angular momentum
 exchange due to the stationary tides. 
In this way the otherwise unspecified time lag $\tau$  may be expressed
in terms of quantities 
determined by the form of viscosity in the planet, the planet's structure,  
and also by the mode of oscillation responding to the tidal field.
Note that in framework of the standard theory of quasi-static tides the similar relations
have been obtained by Zahn (Zahn 1989 and references therein) and by Eggleton, Kiseleva and Hut (1998).

As  can be seen from  equations (\ref{eqno39}) and (\ref{eqno40}),
in the limiting case $\eta \rightarrow \infty$,
the planetary angular velocity  tends to 
evolve faster than its orbital energy with the consequence that
if there are repeated encounters,
the planet achieves a
state of pseudo-synchronisation with $\Omega_{r}=\Omega_{ps}\equiv {33\sqrt 2\over 40\eta}\Omega_{*} \ll \Omega_{*}$, 
at which point, exchange of
angular momentum between the orbit and the planet  ceases (Hut 1981).
For $\Omega_{r}=\Omega_{ps}$ the energy gain
follows from equation (\ref{eqno39}) as    
\be \Delta E_{st}=\beta_{3}{\tilde \gamma 
\tilde Q^{2}\over \tilde \omega_{00}^{4}}{E_{pl}\over \eta^{5}}, \label{eqno42} \ee   
where $\beta_{3}={41\over 195}\beta_{1}\approx 6.64$.
We comment that the angular velocity $\Omega_{ps}$  roughly corresponds
to the orbital angular velocity at closest approach. In the next Section
we use equation (\ref{eqno42})  to estimate the importance of 
stationary tides for the orbital evolution of the planet.

\subsection{Dynamic Tides and the Tidal Excitation of Normal Modes}

When the tidal encounter with the central star occurs on a time-scale which is long but not
extremely long  compared
to the internal dynamical time the planet can be left with a normal mode excited
at some non-negligible amplitude. There will be energy and angular momentum associated with this
mode that is transferred from the orbit. In contrast to the situation for
quasi-static tides the effect does not depend on viscosity. 
We refer to it as the non-stationary
or dynamic contribution to the energy and angular momentum transfer.  

In order to calculate
the non-stationary contribution to the energy and angular momentum gain we can neglect
the viscosity term $\propto \gamma$ in equations  (\ref{eqno24} - \ref{eqno26})
which may then be written in the form
\be \ddot a_{\pm}+\omega^{2}_{00}a_{\pm}\pm \sigma \dot a_{\mp}=f^{T}_{\pm}, \quad
\ddot a_{0}+\omega^{2}_{00}a_{0}=f^{T}_{0}. \label{eqno43}\ee
We use below the solution to equation (\ref{eqno43}) obtained by the method
of variation of parameters   
correct to first order in $\sigma/\omega_{00}$ in the form 
 \be  a_{\pm}={1\over 2\omega_{00}}\int^{t}_{-\infty}dt^{'} 
\lbrace f_{\pm}(\sin\omega_{\pm}(t-t^{'})+\sin\omega_{\mp}(t-t^{'})) 
  +f_{\mp}(\cos\omega_{\pm}(t-t^{'})-\cos\omega_{\mp}(t-t^{'}))\rbrace, \label{eqno44}\ee
\be a_{0}={1\over \omega_{00}}\int^{t}_{-\infty}dt^{'} f_{0}\sin\omega_{00}(t-t^{'}), \label{eqno45}\ee 
where $\omega_{\pm}=\omega_{00}\pm {\sigma \over 2}$. In equations (\ref{eqno44} - \ref{eqno45})
we assume that the perturbations
of the planet are absent before  the close encounter ($t = -\infty $) 
(the so-called first passage problem). 
General expressions  appropriate to repeated encounters
may  easily obtained from the expressions derived for the first passage
problem (see below).

In the limit $t \rightarrow \infty$  
the coefficients  multiplying $\sin(\omega_{00}(t))$,  $\sin(\omega_{\pm}(t))$,
and $\cos(\omega_{00}(t))$, $\cos(\omega_{\pm}(t))$ in 
equations  (\ref{eqno44} - \ref{eqno45}) 
contain  converging integrals. These contain the information
about the residual  mode excitation after the encounter.  Extending  
the upper limit of integration in these integrals to infinity, we have
\be a_{+}={1\over 2\omega_{00}}(I_{+}(\omega_{+})\sin(\omega_{+}t)+I_{-}(\omega_{-})\sin(\omega_{-}t)), \label{eqno46}\ee
\be a_{-}={1\over 2\omega_{00}}(I_{-}(\omega_{-})
\cos(\omega_{-}t)-I_{+}(\omega_{+})\cos(\omega_{+}t)), \label{eqno47}\ee
and
\be a_{0}={1\over \omega_{00}}I_{0}(\omega_{00})\sin(\omega_{00}t), \label{eqno48}\ee
where
\be I_{\pm}(\omega)=\int^{+\infty}_{-\infty}dt(f_{+}\cos(\omega t)\pm f_{-}\sin(\omega t)), \quad
I_{0}(\omega)=\int^{+\infty}_{-\infty}dtf_{0}\cos(\omega t). \label{eqno49}\ee
These integrals can be represented in the form
\be I_{\pm}=2\sqrt 2\pi {GMQ\over \Omega_{p} D_{min}^{3}}K_{m}, 
\quad I_{0}=2\pi {GMQ\over \Omega_{p} D_{min}^{3}}K_{0}, \label{eqno50}\ee
where $m=\mp 2$ for $(\pm)$, and
\be K_{m}={W_{m}\over 2\pi}\int ^{+\infty}_{-\infty}{d(\Omega_{p}t)\over x(t)^{3}} \cos(\omega t+ m\Phi(t)), \label{eqno51}\ee
and $x(t)=D(t)/D_{min}$. 
The integrals (\ref{eqno51})
have been calculated numerically and approximated analytically by Press and Teukolsky (1977).
Lai (1997) has obtained useful analytical expressions for these 
integrals in the limit $(\omega \eta) \rightarrow \infty$  in the form
\be K_{-2}\approx {2{z}^{3/2}e^{-{2\over 3}z}\over \sqrt{15}}(1-{\sqrt \pi \over 4\sqrt z}),
\quad K_{0}\approx -{z^{1/2}e^{-{2\over 3}z}\over 2\sqrt 10}(1+{\sqrt \pi \over 2\sqrt z}), 
\quad K_{2}\approx \sqrt{({3\over 5z})}{e^{-{2\over 3}z}\over 32}(1-{89\over 48z}), \label{eqno52}\ee
where $z=\sqrt 2\omega \eta$.  
In that limit we have $|K_{-2}(z)| > |K_{0}(z)| > |K_{2}(z)|$

Substituting equations (\ref{eqno46} - \ref{eqno47}) into equation (\ref{eqno26b}),
we obtain the gain of angular momentum after ``the first''  periastron passage
\be \Delta L_{ns}={1\over 2\omega_{00}} 
\lbrace I_{+}^{2}(\omega_{+})-I_{-}^{2}(\omega_{-})\rbrace. \label{eqno53}\ee  
Analogously substituting equations (\ref{eqno46} - \ref{eqno48}) into 
equation (\ref{eqno26a}), we obtain the expression for the gain of energy
\be \Delta E_{ns}={I_{0}^{2}(\omega_{00})\over 2}+{1\over 4}((1+{\sigma \over 2\omega_{00}})I^{2}_{+}(\omega_{+})+
(1-{\sigma \over 2\omega_{00}})I^{2}_{-}(\omega_{-})) 
\approx {I_{0}^{2}(\omega_{00})\over 2}+
{1\over 4}(I^{2}_{+}(\omega_{+})+I^{2}_{-}(\omega_{-})), \label{eqno54}\ee
where in the last equality we distinguish between $\omega_{\pm}$ and $\omega_{00}$ only in
the factors depending on $\omega$ exponentially.  

In contrast to related expressions calculated in the approximation of   quasi-static  tides,
the expressions (\ref{eqno53} - \ref{eqno54}) give the energy and angular momentum gain  
associated with some particular (in our case, fundamental)
mode of pulsation. 
The transfer of the energy and angular momentum from the mode to the planet proceeds
during some dissipation time $t_{diss}=\gamma_{00}^{-1}$ which could be much larger than the orbital period
of the planet. 
Note that, in general,  the mode damping rate $\gamma_{00}$ 
is different from the damping rate $\gamma$ 
introduced  for  quasi-static tides.   

The expressions  
(\ref{eqno53} - \ref{eqno54}) 
can be significantly simplified. For a non-rotating planet ($\Omega_{r}=0$) and
large values of $\eta$, we can neglect the contribution of $(-)$ and $(0)$ terms, and obtain from
equations (\ref{eqno50}), (\ref{eqno52}) and  (\ref{eqno54})
\be \Delta E_{ns}(\Omega_{r}=0)\approx {I^{2}_{+}(\omega_{00})\over 4}\approx {16\sqrt 2\over 15}\tilde \omega_{00}^{3}
\tilde Q^{2}\eta e^{-{4\sqrt 2\over 3}(\tilde \omega_{00} \eta )}E_{pl}, \label{eqno55}\ee
and $\Delta L_{ns}\approx {I_{+}^{2}(\omega_{00})\over 2\omega_{00}} \approx
{2\Delta E_{ns}\over \omega_{00}}$.
Of course, these expressions follow
directly from the general expression for the energy gain provided by PT and the general relation between
the mode energy and angular momentum for $m=-2$ (e.g. Friedman and Schutz 1978).

In the general case of a rotating planet, we can substitute the
asymptotic expressions (\ref{eqno52}) in equation (\ref{eqno53}) and obtain
\be \Delta L_{ns} \approx {32\sqrt 2\over 15}{(\tilde \omega_{00}
\tilde Q)}^{2}\eta e^{-{4\sqrt 2\over 3}(\tilde \omega_{+} \eta )}
(1-{9\over 2^{14}{(\tilde \omega_{00}\eta)}^{4}}e^{{4\sqrt 2\over 3} (\tilde \sigma \eta)})L_{pl}, \label{eqno56}\ee
where $\tilde \sigma =\sigma /\Omega_{*}$.

It follows from this expression that the non-stationary
tides can spin up the planet up to the angular velocity  which when
inserted into the right hand side of equation (\ref{eqno56})
gives no additional increase in planet angular momentum or $\Delta L_{ns} =0.$ 
It is readily seen that this angular velocity is given by 
\be \tilde \Omega_{crit}\equiv {\tilde \sigma_{crit}\over 4\beta}={3\over 4\sqrt 2 \beta \eta}
\ln{(({4\over 9})^{1/4}8\tilde \omega_{00} \eta)}
\approx {0.53\over \beta \eta}\ln{(6.53\tilde \omega_{00} \eta)}, \label{eqno57}\ee 
where we  have used  equation (\ref{eqno16}) and the definition of $\sigma$. 

Typically $\beta \approx 0.5$, $\tilde \omega_{00}\sim 1$ and we have 
\be \Omega_{crit}\approx {2+\ln{\eta}\over \eta}\Omega_{*}. \label{eqno57ab}\ee 
It is interesting to note that for distant encounters
$\tilde \omega_{00} \eta$ may be much larger than unity. This has the effect 
that, in contrast to the situation with quasi-static tides,  the critical
angular velocity may easily exceed the orbital angular velocity
at periastron.    For closer encounters with 
sufficiently small values of $\eta$ (but still larger than one)
the critical angular velocity 
can even  be of the order of the characteristic frequency $\Omega_{*}.$ 
This comes about because in simple terms, the pattern speed of the excited normal mode
is also characteristically $\Omega_{*}$
and the star has to rotate by an amount approaching this before
the mode excitation process is significantly affected.

Thus in principle, the non-stationary tides might lead to rotational break-up
of the planet at values of periastron beyond the Roche limit.
However, our approximations are
clearly invalid at such high angular velocities. 

Substituting equation (\ref{eqno52}) into equation  (\ref{eqno54}), we have the similar expression for the energy gain
\be \Delta  E_{ns}\approx {16\sqrt 2\over 15}\tilde \omega_{00}^{3}
\tilde Q^{2}\eta e^{-{4\sqrt 2\over 3}(\tilde \omega_{+} \eta)}
(1+{3\over 2^{6}{(\tilde \omega_{00}\eta)}^{2}}e^{{2\sqrt 2\over 3}(\tilde \sigma \eta)}+
{9\over 2^{14}
{(\tilde \omega_{00}\eta)}^{4}}e^{{4\sqrt 2\over 3} (\tilde \sigma \eta)})E_{pl}. \label{eqno58}\ee 
 
It is easy to see that the energy gain attain its minimal value at $\Omega=\Omega_{crit}$.
Taking into account that the expression in the round brackets in equation (\ref{eqno58}) is equal to $4$ for $\Omega=\Omega_{crit}$,
we have
\be \Delta E_{min}=\Delta E_{ns}(\Omega=\Omega_{crit})=
{3\over 32}{\Delta E_{0}\over {(\omega_{f}\eta)}^{2}}={1\over 5\sqrt 2}{\tilde \omega_{f}
\tilde Q^{2}\over \eta} e^{-{4\sqrt 2\over 3}(\tilde \omega_{00} \eta )}E_{pl}.  \label{eqno58a}\ee 

\subsection{Energy and angular momentum gain  
resulting from multiple encounters}

A planet in a highly eccentric orbit will undergo multiple periastron
passages each of which is associated with an energy and angular momentum
exchange. If dissipative processes are sufficiently weak, normal modes
remain excited between periastron passages with the consequence
that successive passages are not independent. We here analyse
such a situation and find that for highly elongated
orbits there may be a stochastic build up of normal mode energy
at the expense of orbital energy and provide an estimate
of when this will occur.

As pointed out
by several authors (PT, Kochanek 1992, Mardling 1995 a,b)
the amount of energy and angular momentum
gained
(or released) during a  close approach
by some particular mode of oscillation
depends on the mode amplitude and the phase of 
oscillation at the moment of periastron passage.
Let us consider this effect in more detail. 
As indicated above,  we assume that the mode damping time $\sim \gamma_{f}^{-1}$
is much larger than the orbital period of the planet
\be P_{orb}=\pi \sqrt{{GM\over 2E^{3}_{orb}}} \label{eqno60}\ee
and thus  neglect the effect of dissipation.
Here $E_{orb}={GM\over 2a_{orb}}$ is the specific binding energy
of the planet's orbit (from now on  called  the orbital energy)
and $a_{orb}$ is the 
semi-major axis. 

We  assume that the tidal interaction
occurs impulsively. The energy and angular momentum exchanges
are assumed to occur  
at the moment of periastron passage and at any other time 
the mode evolution is described by solutions 
of equations (26-28) without  external forcing terms 
in the form 
\be a_{+}=B_{+}(j)\sin(\omega_{+}t+\psi_{+}(j))+B_{-}(j)
\sin(\omega_{-}t(j)+\psi_{-}(j)), \label{eqno61}\ee 
\be a_{-}=B_{-}(j)\cos(\omega_{-}t+\psi_{-}(j))-
B_{+}(j)\cos(\omega_{+}t+\psi_{+}(j)), \label{eqno62}\ee 
\be a_{0}=\sqrt 2B_{0}(j)\sin(\omega_{00}t+\psi_{0}(j)). \label{eqno63}\ee
Here $B_{i}(j)$, $\psi_{i}(j)$ are the mode amplitudes and phases
after $(j-1)$ periastron passages and we have
introduced the index $(i)$ which can be one  of $(+,-,0)$. 

The  phases have been adjusted 
to make the 
time $t=0$ correspond to the $j$th periastron  passage.
To take into account the interaction at the $j$th periastron passage
we appropriately  add in  the expressions (\ref{eqno46}-\ref{eqno48})
to (\ref{eqno61}-\ref{eqno63}).

The result can also be represented in the generic form 
(\ref{eqno61}-\ref{eqno63})
but with new amplitudes $B_{i}(j+1)$
and phases $\psi_{i}(j+1)$ given by 

\be B_{i}(j+1)=\sqrt {B^{2}_{i}(j)+b_{i}^{2}+2B_{i}(j)b_{i}\cos(\psi_{i}(j))},
 \label{eqno64} \ee 
\be\psi_{i}(j+1)=\arctan ({B_{i}(j)\sin(\psi_{i}(j))\over B_{i}(j)\cos(\psi_{i}(j))+b_{i}})+\phi_{i}(j), \label{eqno65} \ee
where $b_{i}={I_{i}(\omega_{i}) \over 2\omega_{00}}$ for $i=(+,-)$
and $b_{0}={I_{0}(\omega_{00}) \over \sqrt 2\omega_{00}}$ (see also Mardling 1995 a, Mardling $\&$ Aarseth 2001). 
The additional phase $\phi_{i}(j)=\omega_{i}P_{orb}(j)$
being the normal mode phase shift during the orbital period
immediately after the $j$th periastron passage 
is added in to make the time $t=0$ now correspond to the $j+1$th
periastron passage.
We assign the index $(j)$ to the orbital period stressing the fact that the period can change with time.
Obviously, the first term in (\ref{eqno65}) $\sim \psi_{i}(j)$ when $B_{i}(j) \gg b_{i}$.

The expressions for the mode energy and angular momentum just
before  the $j$th 
periastron
passage follow from equations (29-30)
together with  equations (\ref{eqno61}-\ref{eqno63}) as 
\be E_{m}(j)=\sum_{i}{(\omega_{00}^{2}+\omega_{i}^{2})\over 2}B^{2}_{i}(j)
\approx \omega_{00}^{2} \sum_{i}B^{2}_{i}(j), \label{eqno66}\ee
\be L_{m}(j)=2\omega_{00}(B_{+}^{2}(j)-
B_{-}^{2}(j)). \label{eqno67}\ee
The gain of energy and angular momentum after the  $j$th
periastron  passage  follows from equations (\ref{eqno64}), 
(\ref{eqno66}) and  (\ref{eqno67}) as 
\be \Delta E_{m}(j)=\omega_{00}^{2}
\sum_{i}\left((b_{i}^{2}+2b_{i}B_{i}(j)\cos(\psi_{i}(j)))\right),
 \label{eqno68}\ee
\be \Delta L_{m}(j)=2\omega_{00}
\lbrace b_{+}^{2}+2b_{+}B_{+}(j)\cos(\phi_{+}(j))-
(b_{-}^{2}+2b_{-}B_{-}(j)\cos(\phi_{-}(j)))\rbrace . \label{eqno69}\ee
 According to equations 
(\ref{eqno68}) and  (\ref{eqno69}),
the energy and angular momentum can either increase or decrease
after periastron passage
depending on sign of $cos(\psi_{i}(j))$
and the value of $B_{i}(j)$ (PT; Kochanek 1992; Mardling 1995 a,b).

\subsection{A two dimensional map for successive encounters}
Equations (\ref{eqno64}) and 
(\ref{eqno65}) define an iterative map.
This can be  brought into a simpler form by introducing
the two-dimensional vectors ${\bf x}_{i}(j)$
with components
\be x^{1}_{i}={B_{i}(j)\over b_{i}}\cos(\psi_{i}(j)), 
\quad x^{2}_{i}={B_{i}(j)\over b_{i}}\sin(\psi_{i}(j)). \label{eqno70}\ee
We obtain from equations (\ref{eqno64})  and 
(\ref{eqno65}) 
\be {\bf x}_{i}(j+1)={\bf R}(\phi_{i})({\bf x}_{i}(j)+{\bf e}), 
\label{eqno71}\ee
where ${\bf R}(\phi_{i})$ is a
two-dimensional rotational matrix 
corresponding to a rotation through an angle $-\phi_{i}$,
and ${\bf e}$ is a 
two-dimensional constant vector with components $e^{1}=1$, $e^{2}=0$.
The mode energy is expressed in terms of the new
variables as
\be E_{m}(j)= \Delta E_{ns}\sum_{i}
\beta_{i}{|{\bf x}_{i}|}^{2}, \label{eqno72}\ee
where $\Delta E_{ns}$ is given by equation (\ref{eqno58}),
and the dimensionless coefficients $\beta_{i}={{(\omega_{00}b_{i})}^{2} \over \Delta E_{ns}}$.
Note that $\sum_{i} \beta_{i}=1$.

For a constant period $P_{orb}(j)=P_{orb}$, the angles $\phi_{i}$ 
are constant, and the general solution for the mapping given by
equation (\ref{eqno71})
can be easily found  in the  form

\be {\bf x}_{i}(j)={\bf x}^{st}_{i}+{\bf A}(j\phi_{i}){\bf x}_{0}, 
\label{eqno73}\ee
where the constant vectors ${\bf x}^{st}_{i}$   give
the stationary  point  under the mapping defined by equation (\ref{eqno71})
and are  found from the condition that
${\bf x}_{i}(j+1)={\bf x}_{i}(j).$
They have the components
\be x^{st(1)}_{i}=-{1\over 2}, 
\quad x^{st(2)}_{i}={\sin (\phi_{i})\over 2(1-\cos (\phi_{i}))}.
\label{eqno74}\ee
${\bf A}(j\phi_{i})$ is the two-dimensional rotational matrix 
corresponding to a rotation through an angle $-j\phi_{i}$, and
the constant vectors ${\bf x}^{0}$ are determined by the  initial conditions.
According to
equation (\ref{eqno73}), the vectors ${\bf x}_{i}(j)$  lie on 
closed circles   centred on the stationary points ${\bf x}^{st}$ with 
radii determined by the initial conditions.
Obviously, no persistent transfer of energy from the orbit 
to the planet is possible in that case.

When the orbital period changes, the situation is much more complicated.
It is clear  from equations (\ref{eqno70})  and
(\ref{eqno71})
that only the values of $\phi_{i}(j)$,
$\psi_{i}(j)$ mod $2\pi$ are significant.
Provided that the change of orbital period
from one periastron passage to the next is sufficiently large  (namely
$\Delta P_{orb} > 2\pi /\omega_{00}$),  successive phases $\phi_{i}(j)$
mod $2\pi$, and consequently the angles $\psi_{i}(j)$     behave
in a  complicated manner. These
quantities can be approximately  represented 
as uniformly distributed random variables. With
this assumption we can average over
$\cos(\psi_{i}(j))$ in equations (\ref{eqno64}) and
(74) to obtain
\be <B_{i}(j)^{2}>=b_{i}^{2}j, \quad <E_{m}>=\Delta E_{ns}j. \label{eqno75} \ee

This indicates a secular increase of mode energy on average and thus
may be thought of as a stochastic instability.
Note that contrary to the usual random walk problem, 
in our case dispersion grows as fast as the
average value: $\sqrt {<E_{m}^{2}>-<E_{m}>^{2}} \approx <E_{m}>$. Nevertheless, in our estimates of
orbital evolution of the planet we 
assume that the energy gain is determined by from equations (\ref{eqno75}).

In a standard approach to the problem the 
change of the period is caused by exchange of energy between
the orbit and the mode due to tidal interaction.
Assuming that there is no dissipation and radiation
of energy away from the planet,
the sum of the orbital energy and the mode energy is conserved, and
the Keplerian relation (\ref{eqno60}) 
can be used to express the change of orbital energy in terms of change of the
mode energy and close the set of equations (\ref{eqno71})  and
(\ref{eqno72}).
Mardling (1995a, 1995b) numerically
investigated the problem of evolution of the orbital parameters and the mode energy for two convective
polytropic stars of comparable masses.  It was 
found that systems with a substantial eccentricity
exhibit stochastic behaviour.
The origin of this result is clear from qualitative point of view. Indeed, more eccentric orbits have larger semi-major
axes $a$, and therefore larger fractional 
changes of the orbital period  following from a periastron passage. We have
\be \Delta P_{orb}(j)=-{3\over 2}{P_{orb}\over E_{orb}}{\Delta E_{m}(j)\over m_{pl}}
=-6\pi\sqrt{{a^{5}\over {(GM)}^{3}}}{\Delta E_{m}(j)\over m_{pl}}. \label{eqno76} \ee

For our problem it is important to establish a condition
of onset of  stochastic instability.
To do this we  consider a situation 
where the mode energy is much smaller than the orbital
energy of the planet,  so that 
the orbital  period does not change significantly from  some 'initial' value 
$P_{0}$. In that case we can express the angles $\phi_{i}$ in terms of the mode
energy $E_{m}(j)$ through
\be \phi_{i}(j)=\phi_{i}(0)-\alpha_{i}\tilde E_{m}(j), \label{eqno77} \ee
where $\phi_{i}(0)=\omega_{i}P_{0}$.
The dimensionless quantities $\alpha_{i}$ determine how fast
$\phi_{i}(j)$ changes  with  $E_{m}(j)$, and follow from  
equation (\ref{eqno76}). Thus
\be \alpha_{i}=6\pi{\omega_{i}a^{5/2}\over {(GM)}^{3/2}}{\Delta E_{ns}\over m_{pl}}. \label{eqno78}\ee 
The dimensionless mode energy $\tilde E_{m}(j)$ is defined as
\be \tilde E_{m}(j)=E_{m}(j)/\Delta E_{ns}=\sum_{i}
\beta_{i}{|{\bf x}_{i}|}^{2}. \label{eqno79}\ee
Equations (\ref{eqno71})  and
(\ref{eqno77} -  \ref{eqno79})
form a closed set of equations, which depend parametrically on the values of
$\phi_{i}(0)$ mod $2\pi$, $\alpha_{i}$  and  $\beta_{i}$.
As initial conditions we can use
the result of the first passage: $x_{i}^{1}(0)=1, \quad x_{i}^{2}(0)=0$.

We analyse the evolution of this set of equations for $\Omega_{r}=0$.
In that case we can suppress the index $i$ in all expressions and the problem depends parametrically
only on the values of $\alpha$ and $\phi_{0}$ mod $2\pi$.
For a fixed value of $\phi_{0}$
and sufficiently small values of $\alpha,$ the solution  generated from 
equation (\ref{eqno71})
is attracted to a
stationary point near to the one that exists for $\alpha =0$
and which satisfies equation (\ref{eqno74}) (see figure 1).
We comment that the map is not area preserving in general
and that for $\alpha \ne 0$, there are an infinite number of fixed points
that have angles $\phi$ 
that  can be found from the
equation
\be \phi = \phi(0) - {\alpha \over 2(1-\cos(\phi))}. \ee
\begin{figure}
\vspace{8cm}\includegraphics{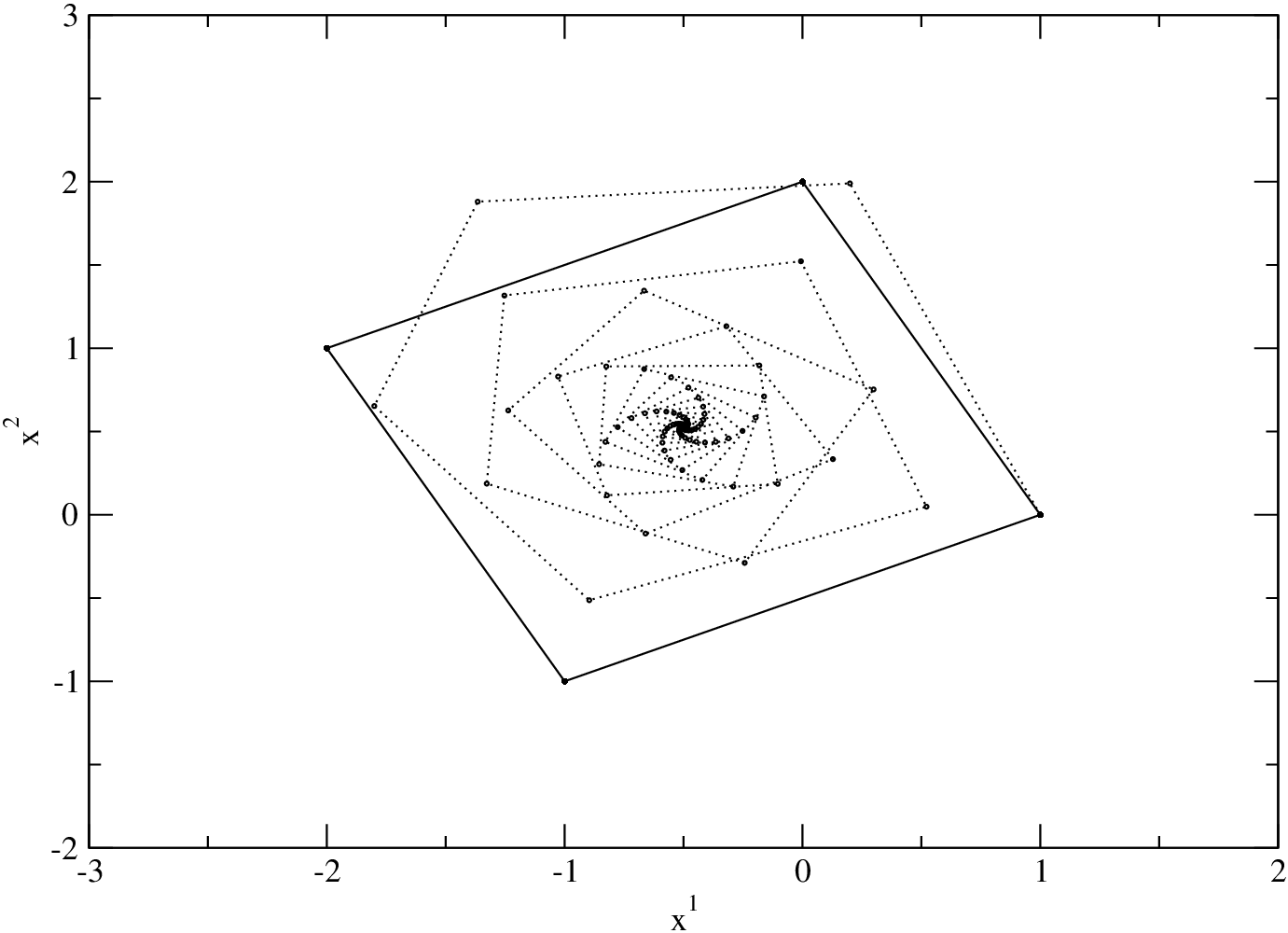}
\caption{The evolution of the mapping (\ref{eqno71})
is shown in the $(x^{1},x^{2})$ plane for
$\alpha=0$ and $\alpha=0.1$.
Circles correspond to positions of the vector ${\bf x}(j)$ for 
 different $j$. Circles connected by
the solid and dotted lines represent the cases
$\alpha=0$, $\alpha=0.1$, respectively.
The angle $\phi_{0}=\pi/2$.
Both curves start at the initial value of ${\bf x}(0)$ with coordinates:
$x_{i}^{1}(0)=1, \quad x_{i}^{2}(0)=0$. The closed curve corresponding to $\alpha=0$ is a parallelogram
where successive values of ${\bf x}(j)$ are situated at vertexes. The curve corresponding to $\alpha=0.1$
is a spiral which ultimately approaches a 
stationary point with coordinates defined by equation (\ref{eqno74}).}
\end{figure}
When $\alpha$ exceeds some value $\alpha_{*}$, this 
stationary  point described above  becomes unstable, and
the solution exhibits a complicated behaviour, but
stochastic growth is not observed (see figure 2).
\begin{figure}
\vspace{8cm}\includegraphics{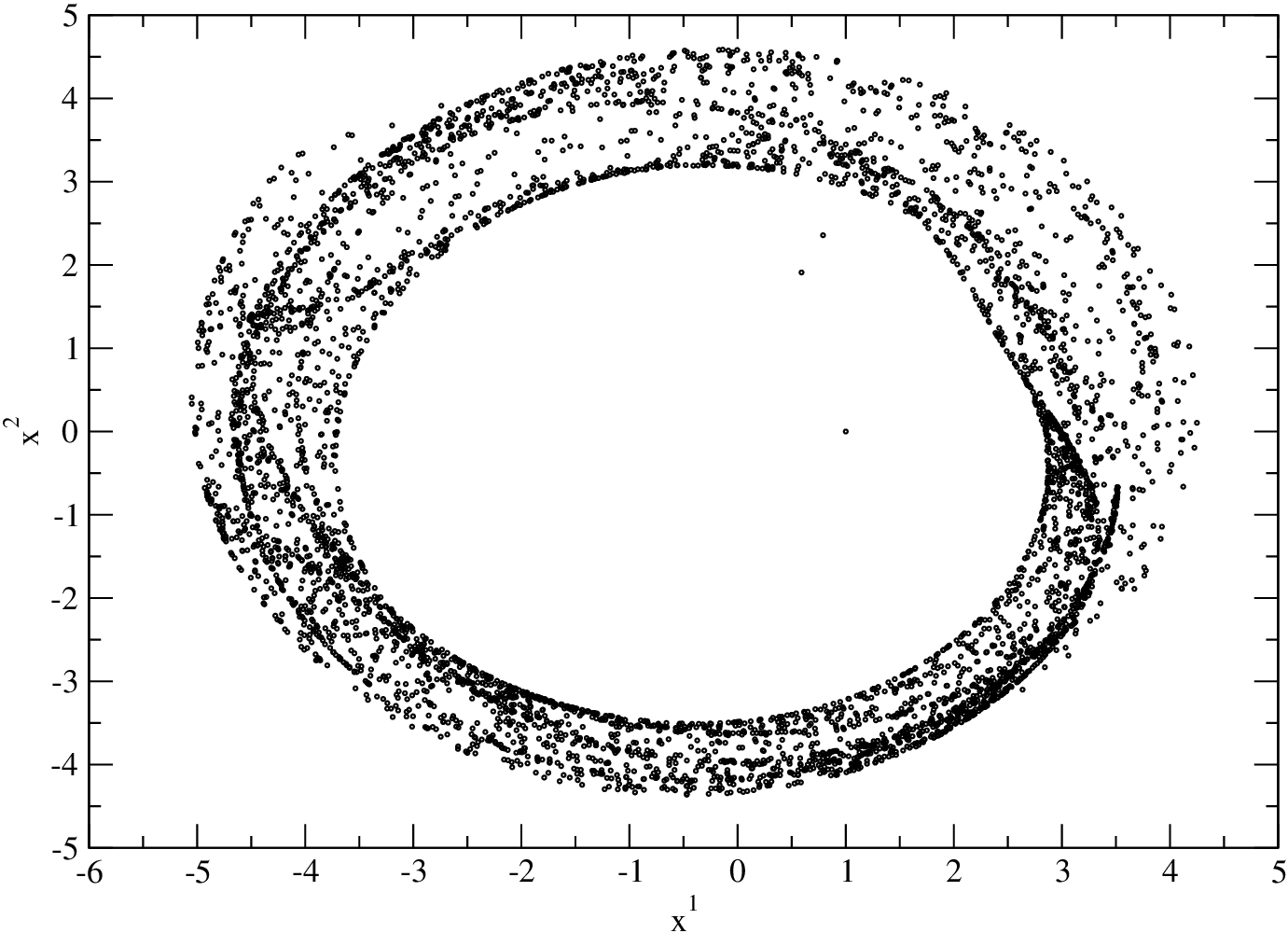}
\caption{Same as figure 1, but with $\alpha=0.3$. We represent here $5000$ point corresponding to
$5000$ iterations of the map (\ref{eqno71}).
The stationary point referred to in the text
is unstable in this case, but
all points are confined to
a small region around it. We have checked for a much
larger number of iterations that $|{\bf x}(j)|$  remains bounded. The case is intermediate between
 attraction to  a fixed point and stochastic behaviour. }
\end{figure}
Finally, when $\alpha$ exceeds some critical value $\alpha_{crit}$, the solution starts to exhibit
stochastic behaviour and the mode energy grows on average (see figure 3).
\begin{figure}
\vspace{8cm}\includegraphics{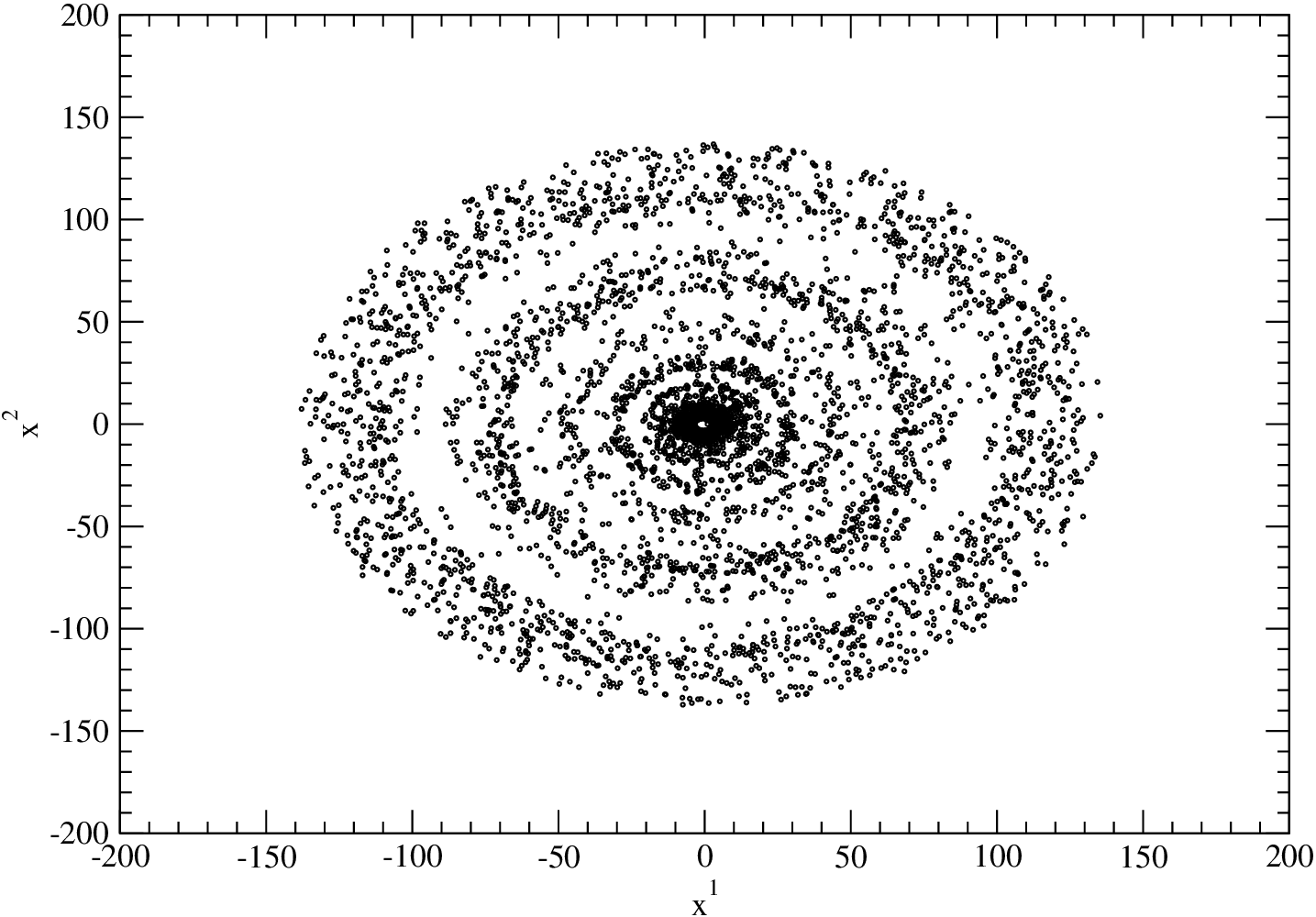}
\caption{Same as Fig. 1  and Fig. 2, but with $\alpha=0.35$. This value of $\alpha$ approximately
corresponds to $\alpha_{crit}$ for $\phi_{0}=\pi/2$ (see Fig. 4).
The evolution can be approximately described as a  jumping of
${\bf x}(j)$ for successive $j$ between different points on circles.
The radius of the circles (proportional to the mode energy)
grows on average.}
\end{figure}
Analytical calculation of
$\alpha_{crit}$ is a very complicated mathematical problem 
so we use a simple numerical
method in order to establish the dependence of $\alpha_{crit}$ on $\phi_{0}$.
Namely, we iterate the map (\ref{eqno71})
for a large number $n$ of iterations ($n=5\cdot 10^{4}$,
$10^{5}$, $1.5\cdot 10^{5}$) and different values of $\alpha$ and $\phi_{0}$
and find the boundary between stochastic and stable solutions. 
We solve the system
starting with $\alpha=0$, and steadily increase $\alpha$
until $\tilde E_{m}(n)$ exceeds the iteration number $n$.
The corresponding value of $\alpha$
is associated with $\alpha_{crit}$. We present the dependence of  $\alpha_{crit}$ on
$\phi_{0}$ in figure 4. One can see from this  plot 
that this value very weakly depends on
the iteration number $n$. For our estimates below we use the averaged value of $\alpha_{crit}$
\be \bar \alpha_{crit}={1\over 2\pi}\int^{2\pi}_{0}d\phi_{0}\alpha_{crit}\approx 0.67.  \label{eqno80aa} \ee

\begin{figure}
\vspace{8cm}\includegraphics{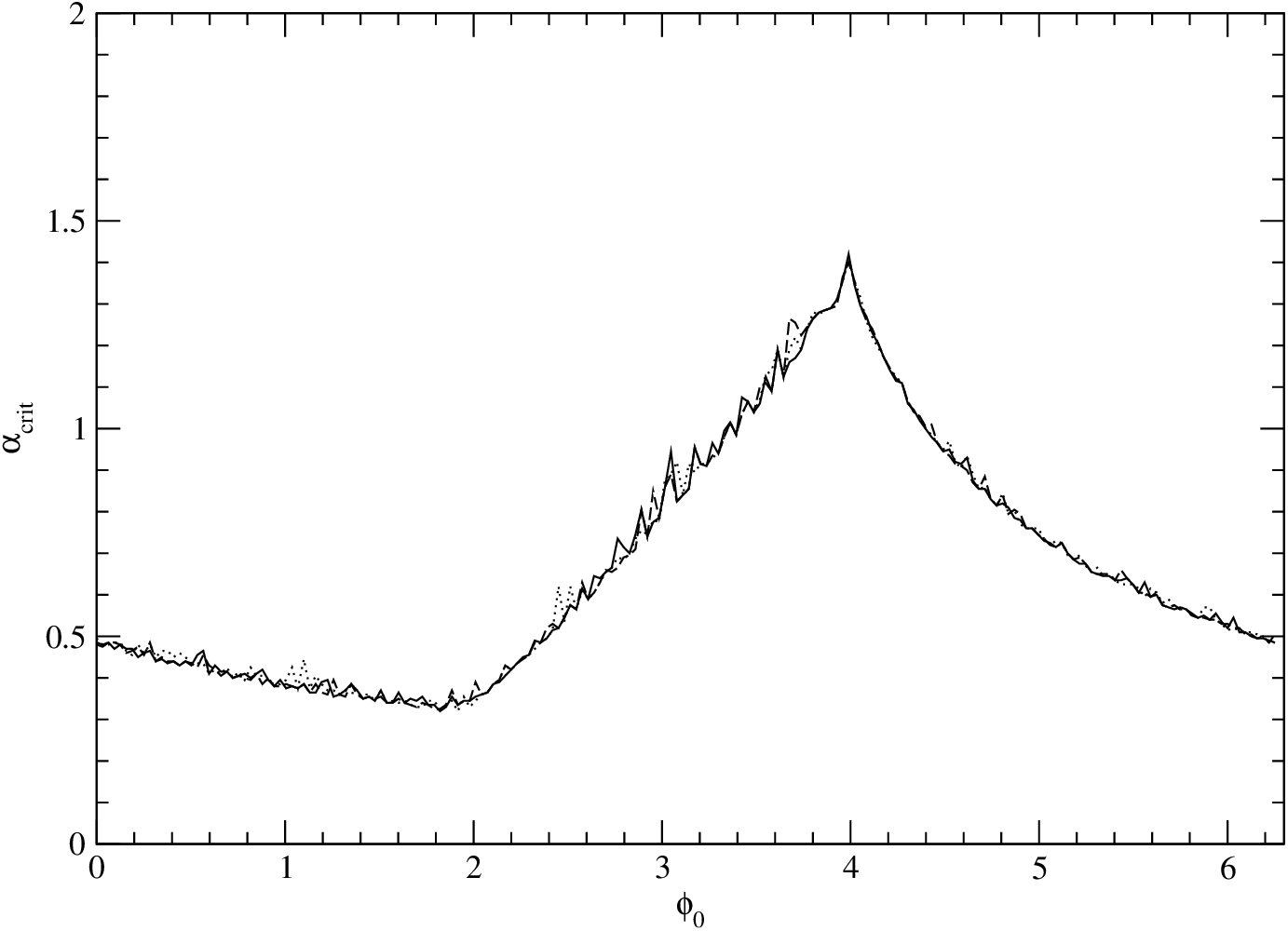}
\caption{The dependence of $\alpha_{crit}$ on $\phi_{0}$ is shown. The solid, dashed,
and dotted lines
correspond to $n=1.5\cdot 10^{5}$, $10^{5}$, $5\cdot 10^{4}$, respectively.}
\end{figure}

One can see from equation (\ref{eqno78}) that $\alpha \propto a^{5/2}$. Therefore, the stochastic instability
always sets in when the semi-major axis $a$ is sufficiently large: $a > a_{st}\equiv a(\alpha=\alpha_{crit})$.
To calculate $a_{st}$ explicitly one should specify a particular model of the planet. We calculate it in 
Section 3.5 below. 

The case of rotating planet can be treated by a similar way, but in this case we have
a large number of parameters defining the map.
The increased number of degrees of
freedom in that case is 
likely to result in decrease of $\alpha_{crit}$ with $\Omega_{r}$.
Therefore, we assume that we can use $\alpha_{crit}(\Omega_{r}=0)$ as an approximate
upper boundary for the rotating case.

\section{Orbital evolution of a massive fully convective planet due to tidal interaction}

In this Section we apply the results of the previous section
to  estimate the effect of tides induced by the central star
on the orbital evolution of a massive planet.
Consider a planetary system consisting of several massive fully  convective planets
orbiting a central star
at typical separation distances $\sim 1-100 Au.$
Suppose that one of these has lost a significant  amount of  angular momentum
(most probably due to interaction with other planets)
and has settled into a highly eccentric orbit. The orbital semi-major axis $a$ would then  be of the order
of 'initial' separation distance $\sim 1-100Au.$ The minimal separation distance $D_{min}$ is related to
the orbital angular momentum per unit of mass $L_{orb}$ through $D_{min}={L^{2}_{orb}\over CM}$ and
could be of the  order of several
stellar radii $R_{*}.$
Situations of this type  occur in simulations of  gravitating interacting planets
(Papaloizou \& Terquem 2001, Adams \& Laughlin 2003).
If close enough approaches to the central star occur,
tidal interaction could become important for determining the subsequent orbital evolution
of the planet. Below we estimate the relative contribution of
stationary and dynamical tides, both exerted by star on  planet
and planet on star on the orbital evolution. In particular
we specify at which separation distances
we can expect tidal effects to become significant.

If we assume that the orbital evolution of the planet is  determined by tides,
while the orbit remains highly eccentric,  the total relative  change of the
orbital angular momentum  is much smaller than  that of the orbital energy.   Thus
we may suppose  the orbital momentum  to be approximately conserved.
That leads to the well known result that the 'final' separation
distance of the planet in the 'final' circular orbit attained at the end of tidal evolution,
$D_{fin}$, is twice as large
as $D_{min}.$
The  shortest observed periods are typically $3$ days for
Jovian mass planets around solar type stars (Butler et al 2002).
Thus $D_{min}$ is directly related to the observed period $P_{obs}$ of the planet
through

\be D_{min}={D_{fin}\over 2}={{(GM)}^{1/3}\over 2}{({P_{obs}\over 2\pi})}^{2/3}=
4.36{(P_{3})^{2/3}}({R_{\odot}\over R_{*}})R_{*}, \label{eqno80} \ee
where $R_{\odot}=7\cdot 10^{10}cm$
is the radius of the Sun, and $P_{3}=P_{obs}/(3days)$. Using equation (\ref{eqno38}) we can also
express the parameter $\eta$ in terms of $P_{3}$:
\be \eta=9.1\eta_{0}{({R_{\odot}\over R_{*}})}^{3/2}P_{3}, \label{eqno82} \ee
where
\be \eta_{0}=\sqrt{{m_{pl}\over M}{R^{3}_{*}\over R^{3}_{pl}}}. \label{eqno83} \ee
For a planet of one Jupiter mass and radius,
and a central solar type star, $\eta_{0}\approx 1$.

\subsection{The planet model}

For a close encounter with the central star occurring
with a given distance of closest approach,
the strength of tidal effects acting in a planet is sensitive
to the magnitude of its radius.
This is because
the energy gain  
through the action of dynamical tides
decreases exponentially with
$\eta$ (see equation(\ref{eqno58})) and $\eta \propto R_{pl}^{-3/2}$
is a monotonically decreasing function of the radius of the planet.
The energy gain  associated with
stationary tides  being $\propto \eta^{-5}$ also decreases rapidly with $\eta.$

In order to determine the radius,
a model for the structure of the planet taking account
its history is needed.
The radius decreases as the planet cools on  a 
thermal time-scale. Accordingly  the model should specify the
planetary structure and radius as a function of age.

A young gravitationally contracting planet will be fully
convective.
If we assume that a turbulent viscosity arising from convection
determines the damping rate $\gamma,$
it will be a function
of the planet radius and luminosity
both of which are functions of time.
In this case too,
estimation of the importance of tidal interaction
requires specification of
a time dependent model of the planet.

\begin{figure}
\vspace{8cm}\includegraphics{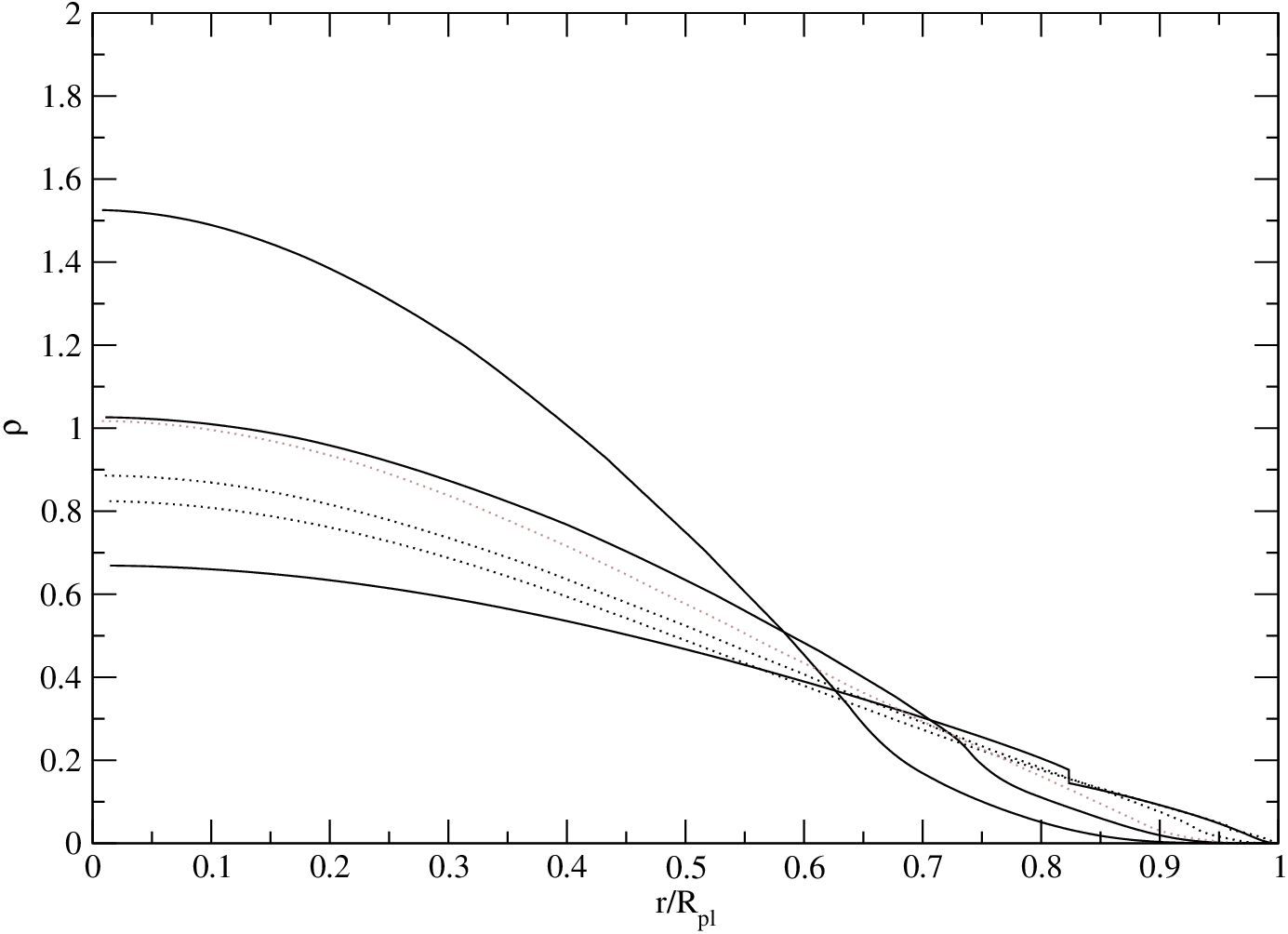}
\caption{The dimensionless density $\tilde \rho$
as a function of the dimensionless radius $r/R_{pl}$. The three solid lines
represent $m_{pl}=1M_{J}$ with $R_{pl}=1R_{J}$, $1.5R_{J}$ and
$2R_{J}$ respectively. The three dotted lines are for
$m_{pl}=5M_{J}$ with $R_{pl}=1.03R_{J}$, $1.54R_{J}$  and
$2.06R_{J}.$ The central density grows with $R_{pl}$ for
both masses.}
\label{Fig5}
\end{figure}

\begin{figure}
\vspace{8cm}\includegraphics{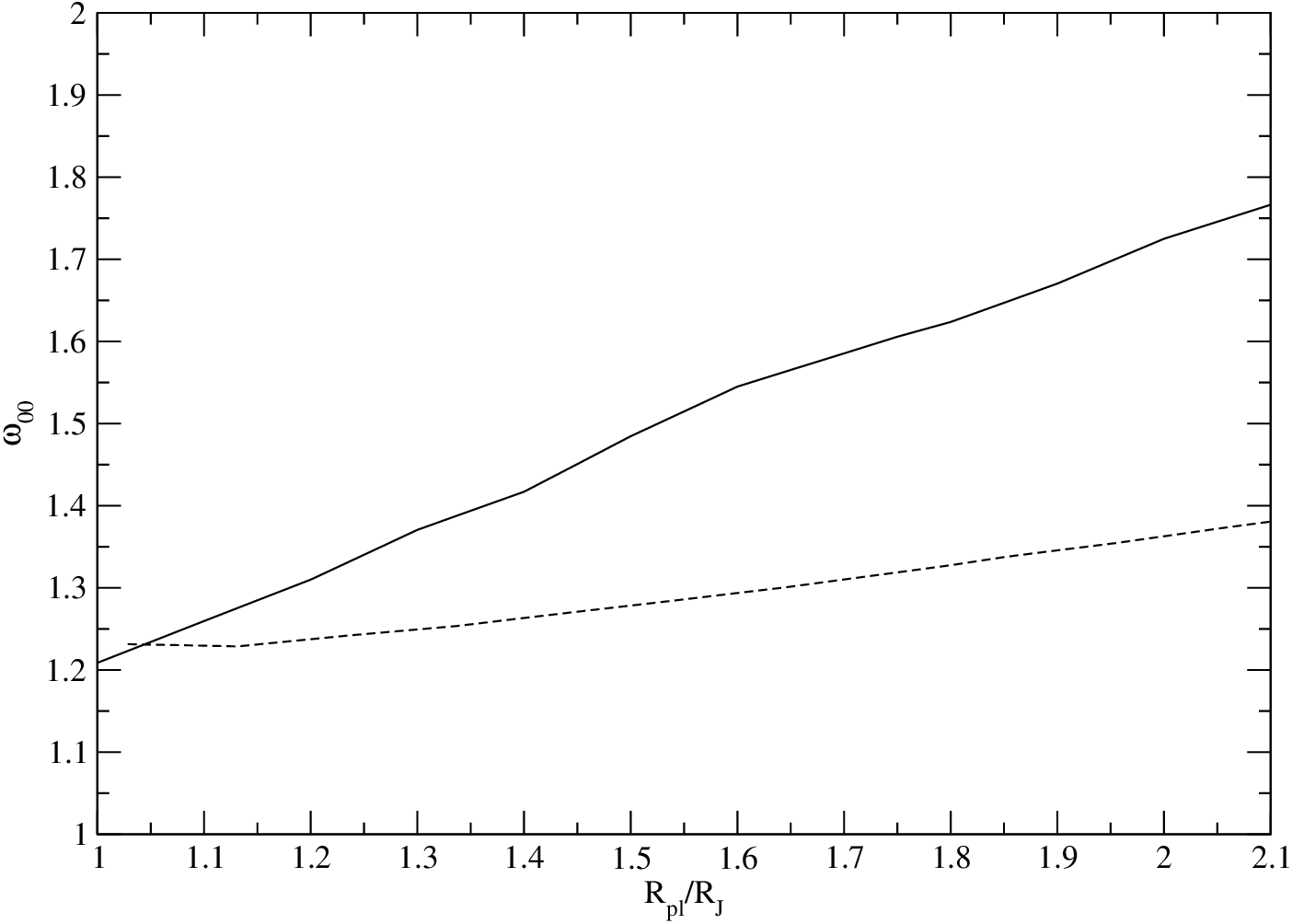}
\caption{The dimensionless frequency $\tilde \omega_{00}$ as a function of the radius of the planet
(in units of the Jovian radius).
The solid and dashed curves correspond
to $m_{pl}=1M_{J}$ and  $5M_{J}$ respectively.}
\label{Fig6}

\end{figure}

\begin{figure}
\vspace{8cm}\includegraphics{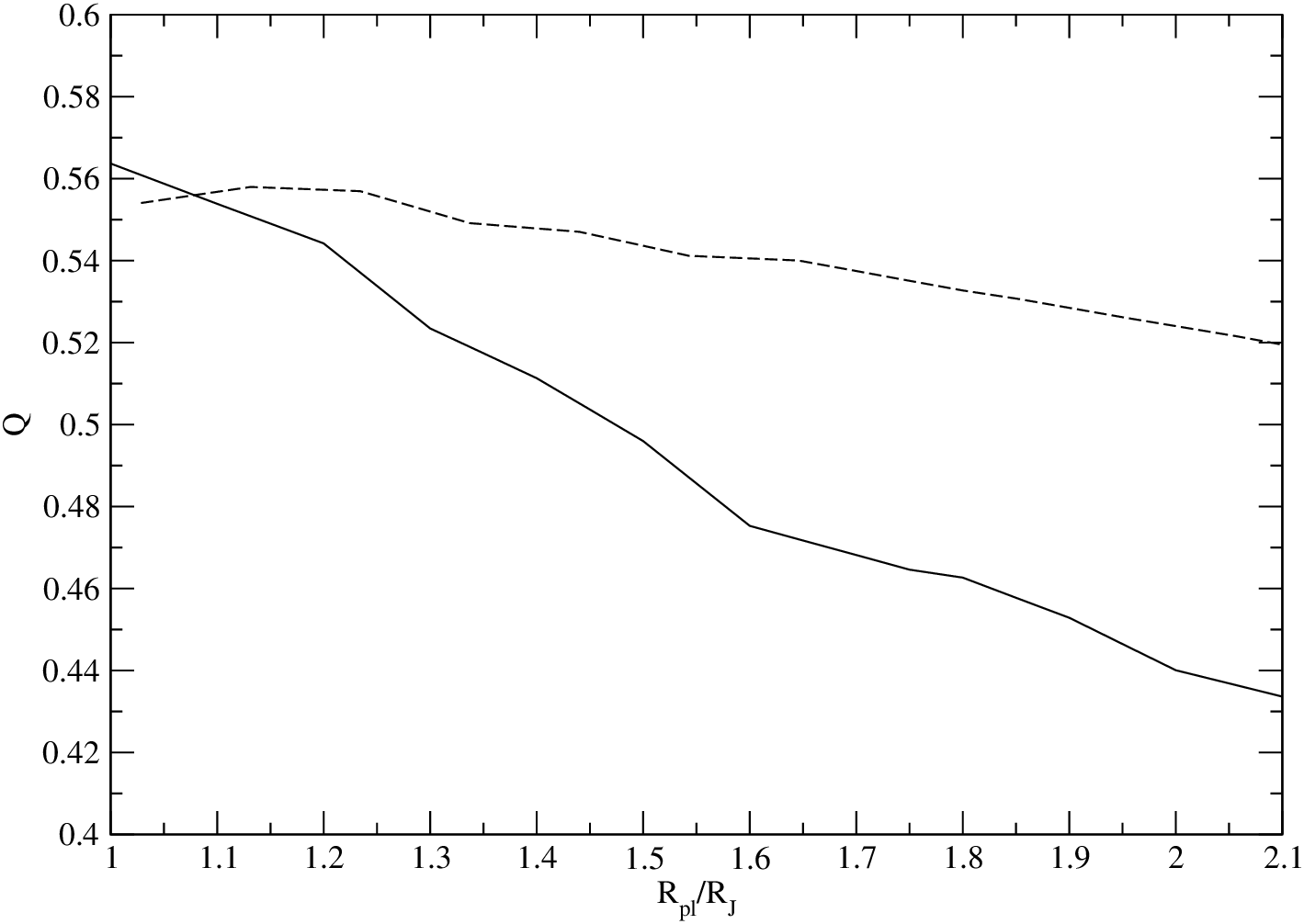}
\caption{The dimensionless overlap integral
$\tilde Q$ as a function of the radius of the planet (in units of the Jovian
radius). The solid and dashed curves correspond to $m_{pl}=1M_{J}$
and $5M_{J}$, respectively.}
\label{Fig7}
\end{figure}

Because convection is extremely efficient,
a fully convective planet can be considered as an isentropic body.
{\footnote{Note that this approximation may break  down
at a boundary between the molecular and metallic phases of
Hydrogen and Helium if the phase transition is of the first kind.
See also Discussion.}}
\noindent For a given composition
isentropic models  of a given mass can be conveniently parametrised
 by the planet radius $R_{pl}.$
We have calculated such
models using the equation of state of Saumon, Chabrier $\&$ Van Horn
(1995) for solar composition neglecting the possible presence of a
central small solid core.
We have considered the masses
$m_{pl}=1M_{J},5M_{J},$ where $M_{J}\approx 1.9\cdot 10^{30}g$
is the mass of Jupiter,
with radii in the range $1R_{J}-2.1R_{J}$, where the
Jovian  radius $R_{J}\approx 7\cdot 10^{9}cm$.
The dimensionless density distribution $\tilde \rho(r)$ as a function of the dimensionless radius $r/R_{pl}$ is shown
in figure \ref{Fig5}.
Note that the models with larger radii are more centrally
condensed, the effect  being
less prominent for $m_{pl}=5M_{J}.$
Also note the sharp change of the density at $r/R_{pl}\approx 0.82$ for
the model with $m_{pl}=1M_{J}$  and $R_{pl}=1R_{J}.$
This is due to the  phase transition between
molecular and metallic Hydrogen that has been assumed to be first order.
 This effect is absent for the models
with larger radii and larger
mass. For these models, the corresponding adiabats have larger entropy
and go above the critical point of
the Saumon, Chabrier $\&$ Van Horn (1995) equation of state.
We calculate the dimensionless frequencies of the
fundamental mode $\tilde \omega_{00}$ and the dimensionless overlap
integrals $\tilde Q$ for these models.
The results of these calculations are presented in figures \ref{Fig6}
and \ref{Fig7}.


The dimensionless
frequencies increase,
 and the dimensionless overlap integrals decrease with $R_{pl}$
these effects   being less
prominent for $m_{pl}=5M_{J}.$
The value of $\tilde \omega_{00}$ for the model with
$m_{pl}=1M_{J}$, $R_{pl}=1R_{J}$ is in a good agreement
with earlier calculations by Vorontsov (1984), Vorontsov et al (1989),
Lee (1993).

The coefficients $\beta$ 
(equation (19)) determining the frequency
splitting due to rotation are close to $0.5$ for all models.
Note that this would be exact for an incompressible
model.

\subsubsection{Evolution of the radius with time}

\begin{figure}
\vspace{8cm}\includegraphics{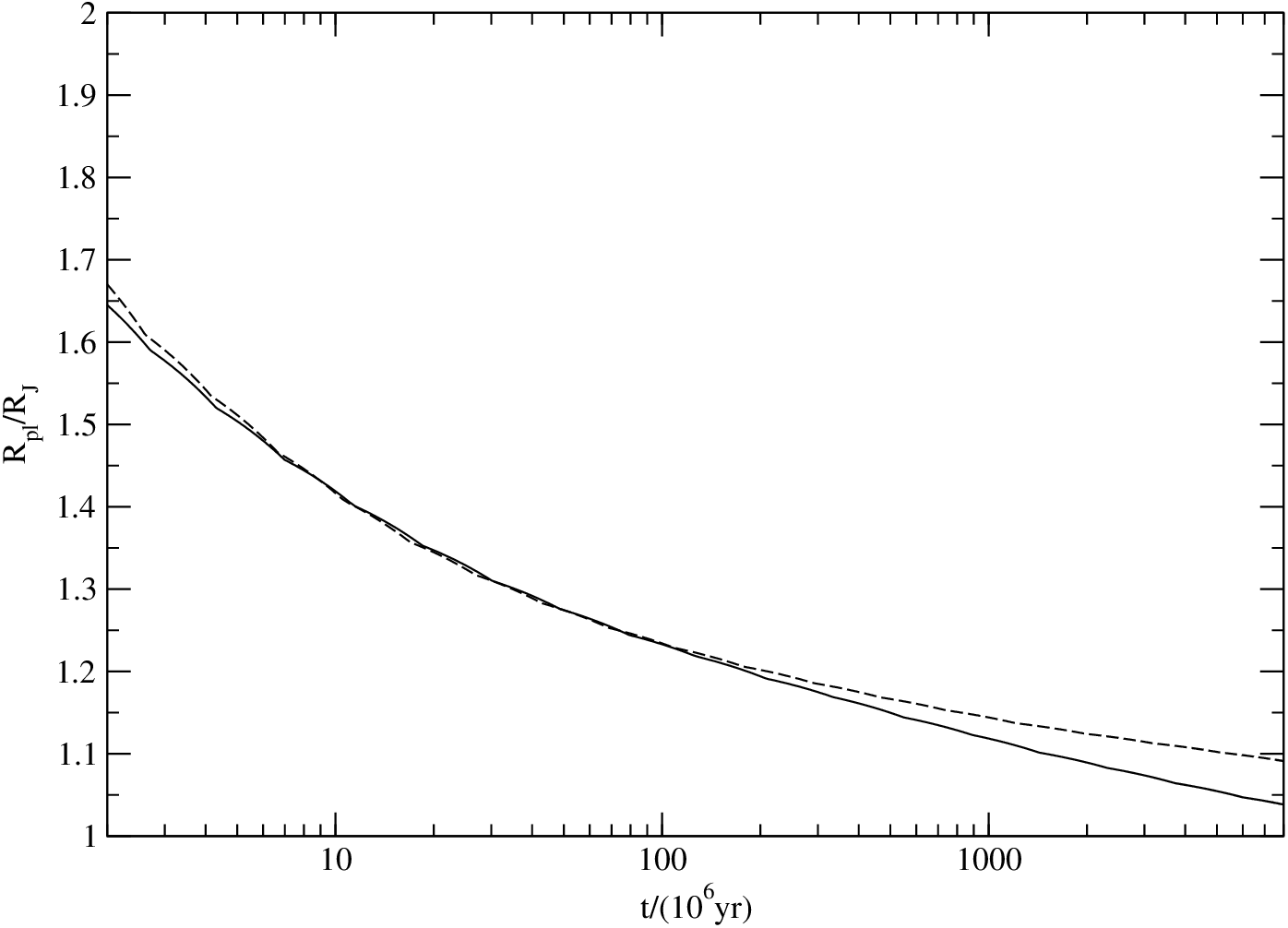}
\caption{The radius of the planet (in units of the Jovian radius)
as a function of time (in units $10^{6}yr$).
The solid and dashed curves correspond to
$m_{pl}=1M_{J}$, $5M_{J}$, respectively.
Note that the dependence of
radius on time is rather similar for both models.}
\label{Fig8}
\end{figure}

\begin{figure}
\vspace{8cm}\includegraphics{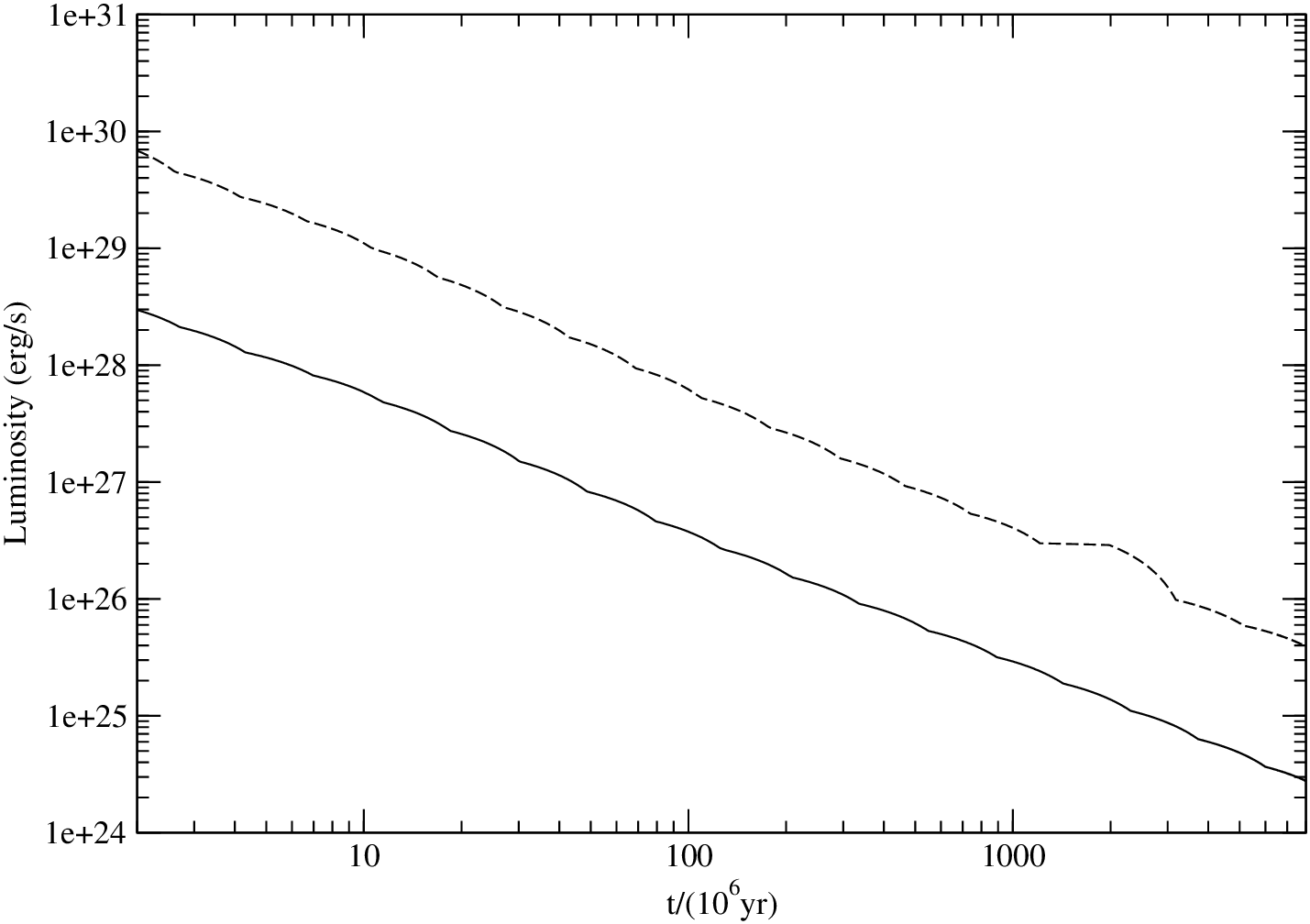}
\caption{The internal luminosity of the planet
(erg/s) as a function of time (in units $10^{6}yr$).
The solid and dashed curves correspond to
$m_{pl}=1M_{J}$  and  $5M_{J}$, respectively. The more massive planet is more than
ten times more luminous.}
\label{Fig9}
\end{figure}

In order to find the dependence of the planetary radius on time,
a model of the planetary  atmosphere is needed. For a given mass and radius,
this is required to determine the luminosity which in turn
determines the cooling rate  as a function of
radius.
This is
a highly non trivial problem given significant
difficulties in calculating molecular opacities for the  low temperatures
and
high densities of interest.
Other complications  may arise from the possible presence of clouds of
different condensed chemical species, etc.
We use the relationship between  radius
and luminosity and hence the planetary  age  calculated by Burrows et al (1997) in the
framework of a non-gray atmospheric model.
The results of
these calculations are shown in figures  \ref{Fig8}
and \ref{Fig9}.
As one can see from figure \ref{Fig8},
the radius of the planet decreases from
$\sim 1.7R_{J}$ at $t=10^{6}yr$ to $\sim R_{J}$ at $t\sim 8\cdot 10^{9}yr$.
Note that this result is close to
that obtained by Graboske et al (1975)
for their model of the  evolution of Jupiter.
We here point out that tidal
heating might significantly delay or even reverse
the cooling of the planet.
This effect is not considered here
as we focus on the initial stages of
tidal evolution and so do not solve a self-consistent problem
for the  thermodynamics of a planet with an energy source
due to tidal heating.

\subsubsection{The convective time scale}

\begin{figure}
\vspace{8cm}\includegraphics{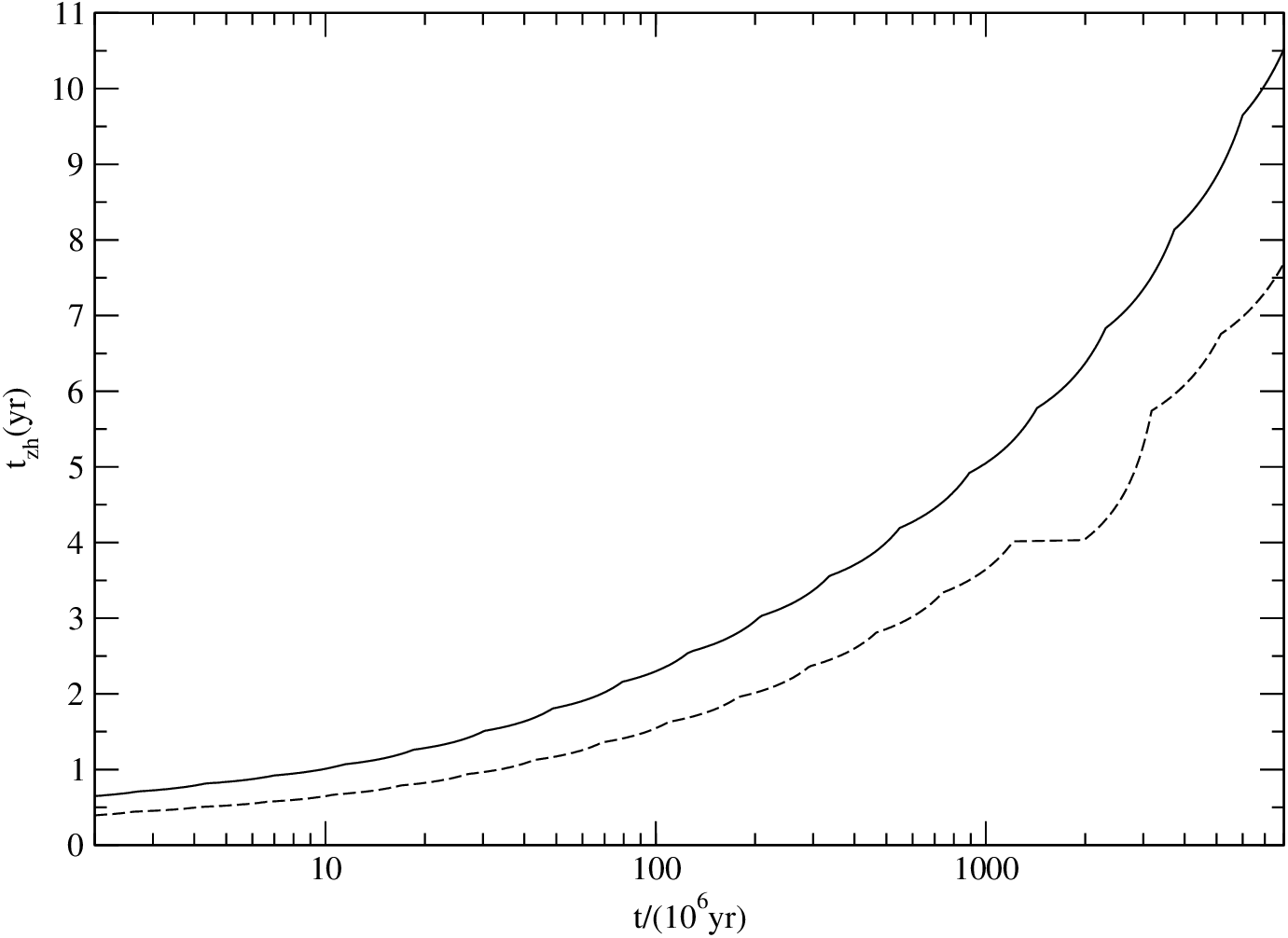}
\caption{The convective time scale
$t_{zh}$ as a function of time (in units $10^{6}yr$).
The solid and dashed curves correspond to $m_{pl}=1M_{J}$  and
$5M_{J}$ respectively. Note that this time scale
is the same order of magnitude for both masses.}
\label{Fig10}
\end{figure}

An important time scale
is the characteristic time  associated with
convective   eddies $t_c$.
These will be of large scale
in the fully convective planetary interior. This characteristic time
has been estimated by Zahn (1977) to be
\be t_c \sim t_{zh}\equiv {\left({m_{pl}R_{pl}^{2}\over 
\dot E_{th}}\right)}^{1/3}, \label{eqno84} \ee
where $\dot E_{th}$ is the energy transport rate due to convection.
This is comparable to the total luminosity.
The dependence of $t_{zh}$ on time is shown in  figure \ref{Fig10}.

One can see from figure \ref{Fig10}
 that $t_{zh}$ grows with time  from    $\sim 0.5 yr$  at an age of
$10^{6}yr$
to $\sim 10yr$ at an age of $10^{9}yr.$
Assuming that dissipation is determined by a
turbulent viscosity that acts like an anomalously  large molecular viscosity,
we can express the dissipation rate
given by equation (20) as
\be \gamma={\lambda(m_{pl}, R_{pl})\over t_{zh}}, \label{eqno85} \ee
where the coefficient
$\lambda(m_{pl}, R_{pl})\sim 10^{-2}-10^{-1}$
for different models of the planet.

In fact the dissipation rate may be much smaller than that given by
equation ({\ref{eqno85}).
This is because as the largest convective eddies have
the size of order of the mixing length, which in the deep interior is on the order
of the local radius,
they can only fully  distribute the energy and momentum
of the hydrodynamical motion induced by the tidal field on the
mixing length scale in a time $\sim t_{c}\gg t_{f}$,
where
$t_{f} \sim 1day$ is the characteristic time associated with  tidal forcing
(Goldreich $\&$ Nicholson 1977).
This leads to a correction factor $\sim {({t_{f}\over t_{c}})}^{n}$
in equation ({\ref{eqno85}).
In the literature it has been suggested that
$n$ may be equal either to $2$
(e.g. Goldreich $\&$ Nicholson 1977) or
to $1$ (Zahn 1989).
Neglecting the contribution of layers close to the surface of the planet,
we  adopt
$t_{c}\sim t_{zh}$, and modify the expression ({\ref{eqno85})
taking into account the correction factor so obtaining
\be \gamma_{m}\sim {({t_{f}\over t_{zh}})}^{n}
({\lambda(m_{pl}, R_{pl})\over t_{zh}}). \label{eqno86} \ee

\subsection{Orbital evolution due to  quasi-static tides}

We show here that  in our approximation,
the time scale for orbital evolution due to  quasi-static tides
of a planet in a highly eccentric orbit
is exceedingly large.
In order to make an estimate of the evolution time scale we  assume the
planet to be rotating at its 'equilibrium'
angular velocity $\Omega_{ps}$ and therefore   evolving at fixed
orbital angular momentum.
This is reasonable on account of the low moment of inertia
associated with the planetary rotation compared to that
of the orbit.
The planet in a highly eccentric orbit can be considered to
have a  sequence of impulsive energy gains  occurring at  periastron passage.
Each energy gain $\Delta E_{st}$ is given by equation (47).
Since the energy gain depends on orbital parameters only through
the dependence of $\eta$ on the
minimum separation distance
and for highly eccentric orbits the minimum separation distance
is determined by the value of the orbital angular momentum,
the energy gain $\Delta E_{st}$ is constant during the
orbital evolution.

The law of energy conservation  gives
\be {\Delta E_{st}\over P_{orb}}+{GMm_{pl}\over 2a^{2}}
\dot a=0, \label{eqno87} \ee
where $a$ is the semi-major axis,
 and $P_{orb}=2\pi\sqrt{{a^{3}\over GM}}$
is the orbital period (also see
equation (66)). From  equation (\ref{eqno87})
we obtain the evolution equation for the semi-major axis in the form
\be {\dot a\over a}=-{1\over t_{ev}}, \label{eqno88} \ee
with
\be t_{ev}=\pi \left({GMm_{pl}\over a \Delta E_{st}}\right )
\sqrt{{a^{3}\over GM}}. \label{eqno89}\ee
It is convenient to introduce a scale $a_{*}$ defined by the condition
that the planetary internal energy is approximately equal to the orbital energy
in the form
\be a_{*}={M\over m_{pl}}R_{pl}\approx 7\cdot 10^{12}\times
\left({R_{pl}\over R_{J}}\right )\left({MM_{J}\over M_{\odot}m_{pl}}\right)cm. \label{eqno90}\ee
Using this we can  express $t_{ev}$ in the form
\be t_{ev}={E_{pl}\over 2\Delta E_{st}}{\left ({a\over a_{*}}\right )}^{1/2}P_{*},
\label{eqno91}\ee
where
\be P_{*}\equiv P_{orb}(a=a_{*})=
0.32\left({M\over M_{\odot}}\right ){\left({M_{J}R_{pl}
\over m_{pl}R_{J}}\right )}^{3/2}yr.  \label{eqno92}\ee

One can easily estimate from equation (\ref{eqno91})
that $t_{ev}$ is very large. Indeed,
$\Delta E_{st}/E_{pl}$ can  be estimated from equation (47)  to be
\be \Delta E_{st}/E_{pl}\sim
 {\tilde \gamma \over \eta^{5}},  \label{eqno93}\ee
 where we have assumed that $\tilde \omega_{00}$ and $\tilde Q \sim 1$.
Taking into account that $t_{zh}\sim 1-10yr$, and $\lambda \sim 10^{-2}$,
we  find  $\tilde \gamma$   using  equation (\ref{eqno85}) to be
$\tilde \gamma \leqsim 10^{-6}$.
Since $\eta \sim 10$  in our problem,
we  obtain $\Delta E_{st}/E_{pl}\leqsim 10^{-11}$, and thus  
$$t_{ev}\sim 10^{11}{({a\over a_{*}})}^{1/2}{({M_{J}\over m_{pl}})}^{3/2}yr.$$
More accurately, we can take into account the fact that
$\Delta E_{st}$ depends on several quantities which are
 functions of time, making in turn
$t_{ev}$ a function of time also.
We show the dependence of $t_{ev}$ on $t$ in figure \ref{Fig11}
 assuming that $\eta=10$, $a=a_{*}$ and the damping rate is given
by equation (\ref{eqno85}).
\begin{figure}
\vspace{8cm}\includegraphics{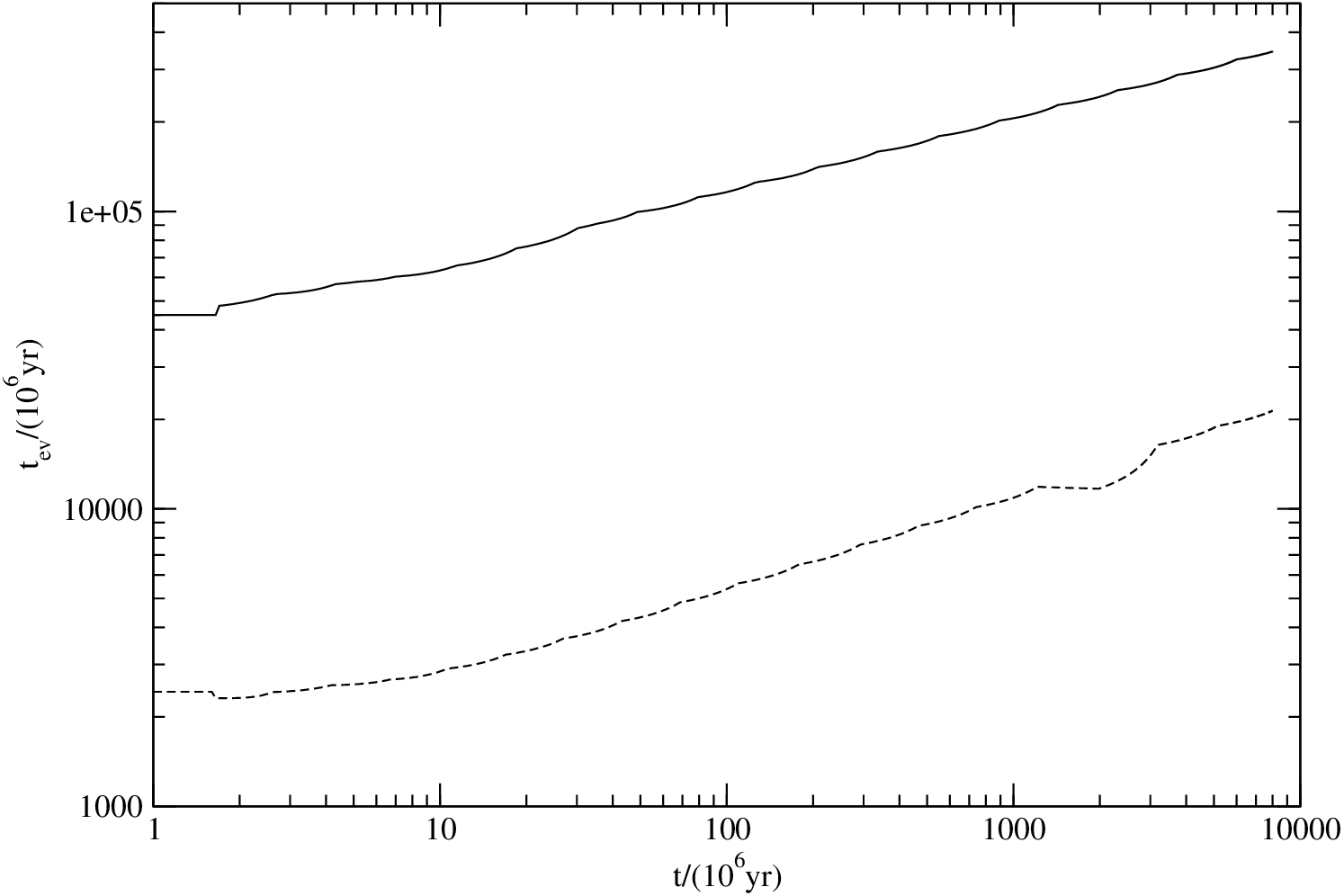}
\caption{The time scale
 of evolution due to quasi-static tides, $t_{ev},$
(in units of  $10^{6}$yr) as a function of time
( in the same units) for $a = a_*$. The solid and dashed curves
 correspond to $m_{pl}=1M_{J}$ and $5M_{J}$ respectively.}
\label{Fig11}
\end{figure}
One can see from  figure \ref{Fig11}
that  the time scale of orbital evolution
due to quasi-static tides is very large, being typically
the order of $\sim 10^{9}-10^{10}yr$ for $m_{pl}=5M_{J}$, and
order of magnitude larger for $m_{pl}=M_{J}$. 
In fact, the evolution time may even be much larger if we take into
account the correction factor to the damping rate (equation (\ref{eqno86})).
Therefore, if we assume that the viscous damping
rate is determined by  convective turbulent viscosity,
the evolution of a planet's orbit due to
quasi-static tides appears to be
impossible for semi-major axes $\geqsim 0.1Au$.

We have also made simple estimates of the evolution due
to  quasi-static tides acting
in a star with a solar-like convective envelope
and found that the rate  of  evolution of a  planet orbit with
orbital parameters typical of our problem
due to tides in the star also seems to be rather weak.

\subsection{Dynamic Tides in  the star versus Dynamic Tides in  the planet}

In order to estimate effects due to
dynamic tides in the star we model it as a
 simple $n=3$ polytrope  and
calculate the energy gain due to dynamic tides,
$\Delta E_{ns*}$ using the expressions  given
by PT. Our results agree well with those of Lee and Ostriker (1986),
but we include the  contribution of a  larger
number of $g$ modes to $\Delta E_{ns*}$ and therefore
obtain a slightly larger value of $\Delta E_{ns*}$ for
large separation distances.
It is important to note that the quantity $\eta_{*}$
associated with tides in the star
is different from that  associated with the planet
(equation (43)) being  given by the expression (PT)
\be \eta_{*}={({D_{min}\over R_{*}})}^{3/2}
\approx 9{({R_{\odot}\over R_{*}})}^{3/2}P_{3}.  \label{eqno94} \ee
In order to show the values of $\Delta E_{ns}$
 and  $\Delta E_{ns*}$ in the same figure,
we use the orbital period $P_{obs}$ as
independent variable.
These quantities are
represented in figure \ref{Fig12}
and figure \ref{Fig13} for $m_{pl}=1M_{J}$ and $m_{pl}=5M_{J}$ respectively.
\begin{figure}
\vspace{8cm}\includegraphics{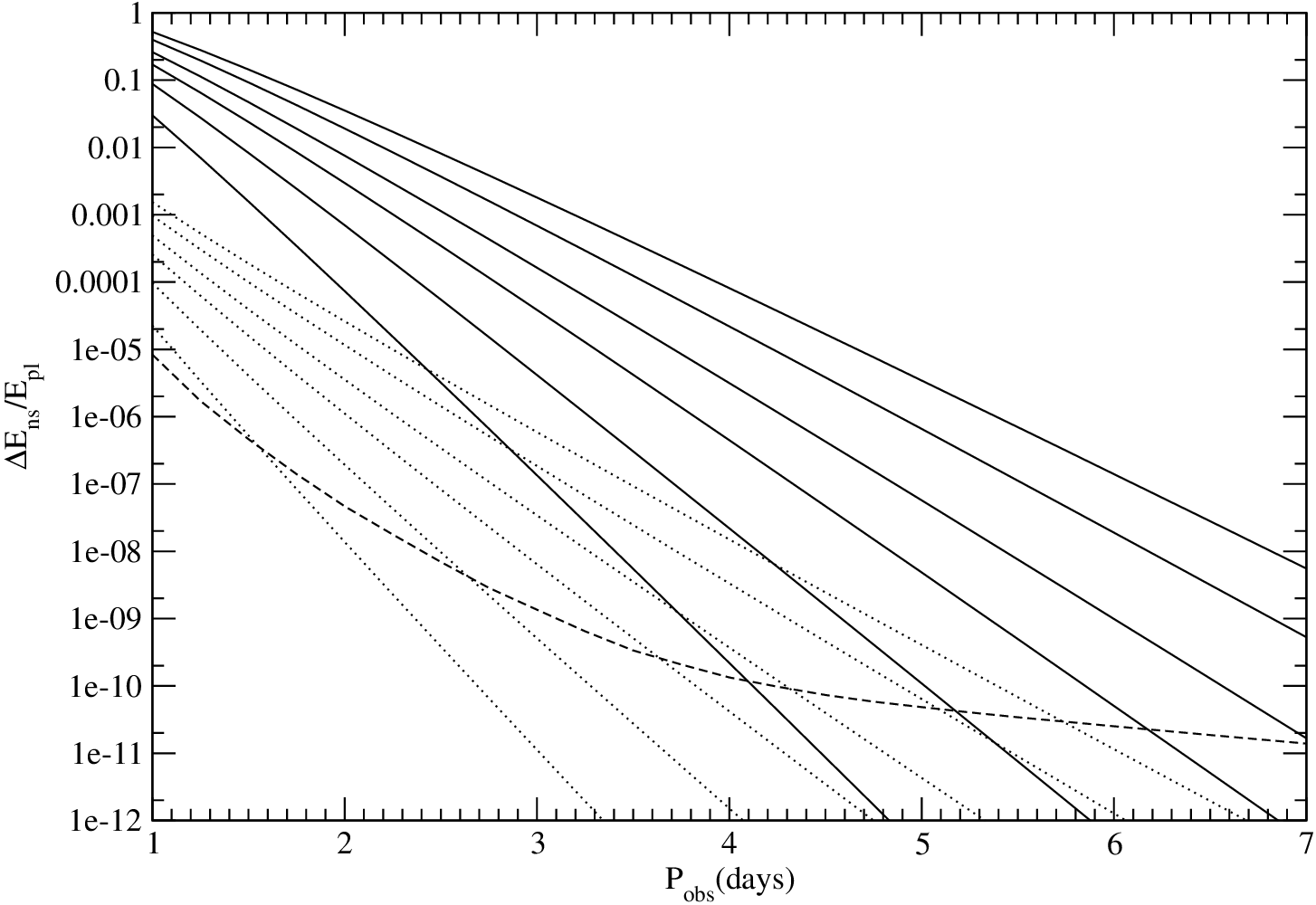}
\caption{The energy gains $\Delta E_{ns}$, $\Delta E_{ns*}$
obtained by the planet (solid and dotted curves)
and by the star (the dashed curve)
after 'the first' periastron fly-by.
The six solid curves correspond to a non-rotating planet
($\Omega_{r}=0$) and
represent the expression
\ref{eqno54}) for the six different values $R_{pl}=1R_{J}$,
$1.2R_{J}$, $1.4R_{J}$, $1.6R_{J}$, $1.8R_{J}$
and $2R_{J}$.
Obviously, the curves corresponding to larger values of $R_{J}$
have larger values of $\Delta E_{ns}$.
The six dotted curves represent the same quantities but calculated
for a planet with  'critical' rotation rate
$\Omega_{r}=\Omega_{crit}$. They are given by the expression
(65).}
\label{Fig12}
\end{figure}
\begin{figure}
\vspace{8cm}\includegraphics{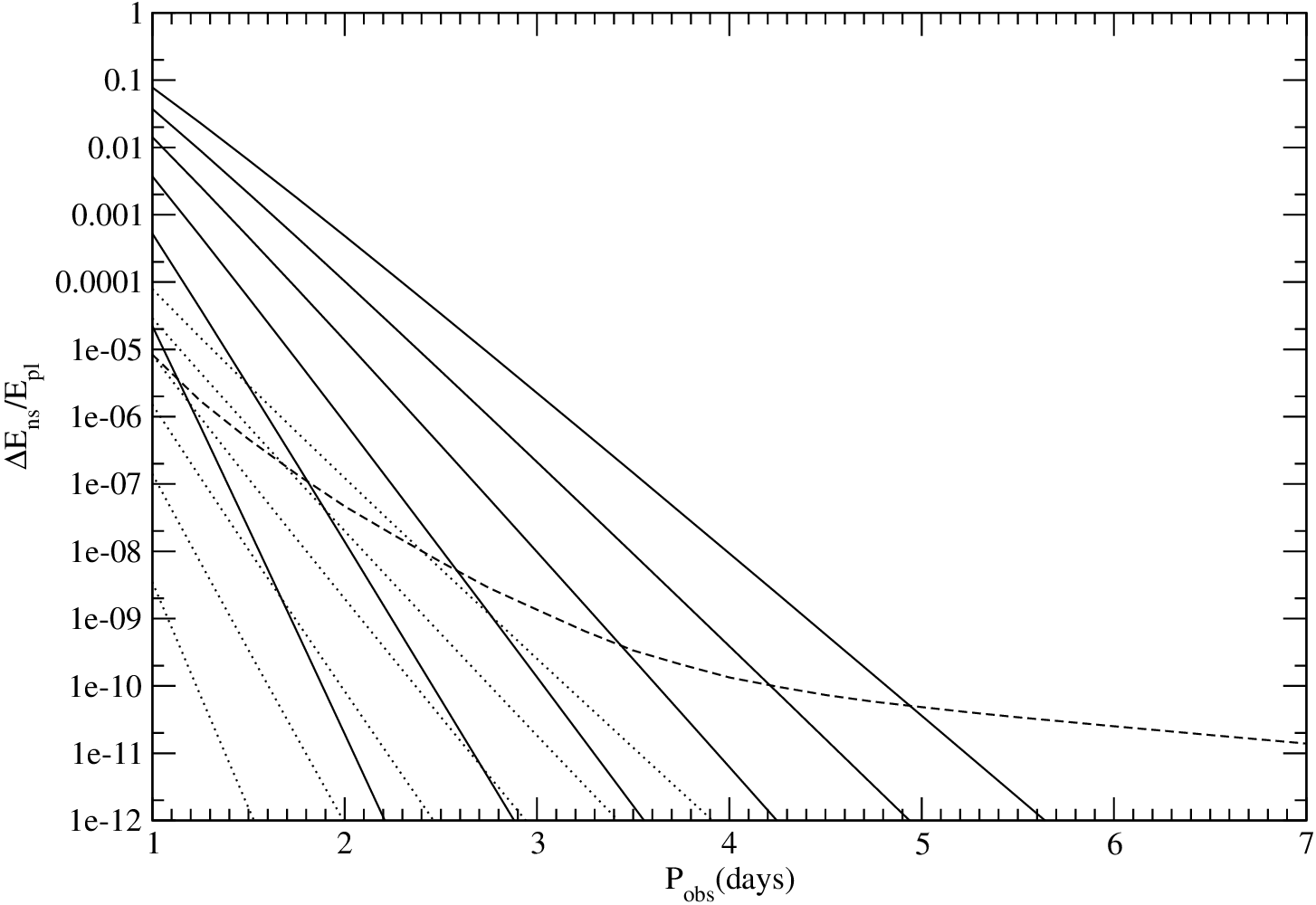}
\caption{Same as \ref{Fig12}, but calculated for $m_{pl}=5M_{J}$.}
\label{Fig13}
\end{figure}
One can see from figure 12
that the energy gain $\Delta E_{ns}$ sharply decreases with $P_{obs}$.
However, in most cases
$\Delta E_{ns} \gg \Delta E_{ns*}$ at $P_{obs}\sim 1day$.
For a non-rotating planet with $m_{pl}=1M_{J}$ tides in the planet
are dominant  for all interesting  values
of $\Delta E_{ns}$: $\Delta E_{ns}/E_{pl} \geqsim 10^{-10}-10^{-9}$.
Even when the planet
rotates at $\Omega=\Omega_{crit}$
(corresponding to minimal possible energy gain for a given $P_{obs}$,
see equation (65)),
the tides in the planet dominate for models with $R_{pl} > 1.4R_{J}$.
The planet with $m_{pl}=5M_{J}$ is more dense,
and the role of tides operating in the planet is less prominent.
In that case the curves corresponding to
'critical' rotation are situated below
the curve corresponding to the tides in the star for almost   the whole
range of
$P_{obs}.$

\subsection{Orbital evolution due to Dynamic Tides}

If we assume that either each impulsive energy transfer to the planet
is dissipated before the next, or that
the condition of the stochastic build-up of mode energy is fulfilled, in either case  we can write
the equation for the
evolution of the semi-major axis due to dynamical tides in a form similar to equation (\ref{eqno88})
\be {\dot a_{10}\over a_{10}}=-{1\over t_{10}(t)\sqrt{a_{10}}}, \label{eqno95} \ee
where $a_{10}=a/10Au.$ The time
\be t_{10}=7.4\cdot 10^{8}\left({MM_{J}\over M_{\odot}m_{pl}}\right)\left ({R_{pl}\over R_{J}} \right )\epsilon_{ns}^{-1}yr,
\label{eqno96} \ee
is the 'local' evolution time for the orbit with semi-major axis $\sim 10Au \gg a_{*}$ and
$\epsilon_{ns}=10^{9}(\Delta E_{ns}+\Delta E_{ns*})/E_{pl}.$  We would like to stress that
if the tidal energy transfer process is in the stochastic regime,
equation (\ref{eqno95}) must be
understood as describing the evolution of an average value of $a$ (see previous Section).

Equation(\ref{eqno95}) can be integrated to give
\be a_{10}=a_{in}\left (1-{1 \over 2\sqrt{a_{in}}}
\int^{t}_{t_{in}}{dt^{'}\over t_{10}(t^{'})}\right )^{2} =  a_{in}\left (1-{\sqrt{a_{cap}\over a_{in}}}\right)^2,
\label{eqno97} \ee
where $a_{in}$ and $t_{in}$ are the 'initial'
values of the semi-major axis (in units  of $10Au$) and the initial  time respectively.
The quantity
\be a_{cap}={1\over 4}\left (\int^{t}_{t_{in}}{dt^{'}\over t_{10}(t^{'})}\right )^{2} \label{eqno98}\ee
defines the typical 'capture' scale.
This is because at any moment of time $t,$ the orbits with initial values of the semi-major
axes $a_{in}<a_{cap}(t)$ have changed their semi-major axes significantly before the time $t$ and
the orbits with $a_{in} > a_{cap}(t)$ have $a \sim a_{in}.$

We can define the evolution time more accurately as the time when the term in brackets
in equation (\ref{eqno97})  is equal
to zero. Thus
\be {1\over 2\sqrt{a_{in}}}\int^{t_{ev}}_{t_{in}}{dt^{'}\over t_{10}(t^{'})}=1. \label{eqno99} \ee
Obviously, this time is a function of $a_{in}$ and $t_{in}$. For a constant value of $t_{10},$
$t_{ev}\approx 2\sqrt{a_{in}}t_{10}.$

The results of calculation of $t_{ev}$ and $a_{cap}$ are presented in  figures \ref{Fig14}-\ref{Fig15}.
For our calculations we assume that
$t_{in}\sim 10^{6}yr$
and note that  results can be easily generalised for other values of
$t_{in}$. In figure \ref{Fig14}
we show  results of calculations  of $t_{ev}$ for the  two  planetary masses  ($m_{pl}=1M_{J},5M_{J}$),
and two values of the angular velocity of the planet ($\Omega_{r}=0,\Omega_{crit}$).
 The orbit of non-rotating $1M_{J}$ planet
can be evolved by dynamical tides in a time $\leqsim 10^{10}yr$ for $P_{obs} < P_{max}\sim 5days$
(see the lower solid curve in figure \ref{Fig14} ).
The orbit of a $1M_{J}$ planet with $\Omega_{r}=\Omega_{crit}$ has
$P_{max}\approx 4days$.  Orbits of non-rotating and rotating
planets with $m_{pl}=5M_{J}$  have $P_{max}\sim 5-6 days.$
In that case the evolution is determined by dynamical tides in the
star for $P_{obs}\geqsim 3days$ (both dashed curves almost coincide in figure \ref{Fig14}
for $P_{obs} \geqsim 3days$).
\begin{figure}
\vspace{8cm}\includegraphics{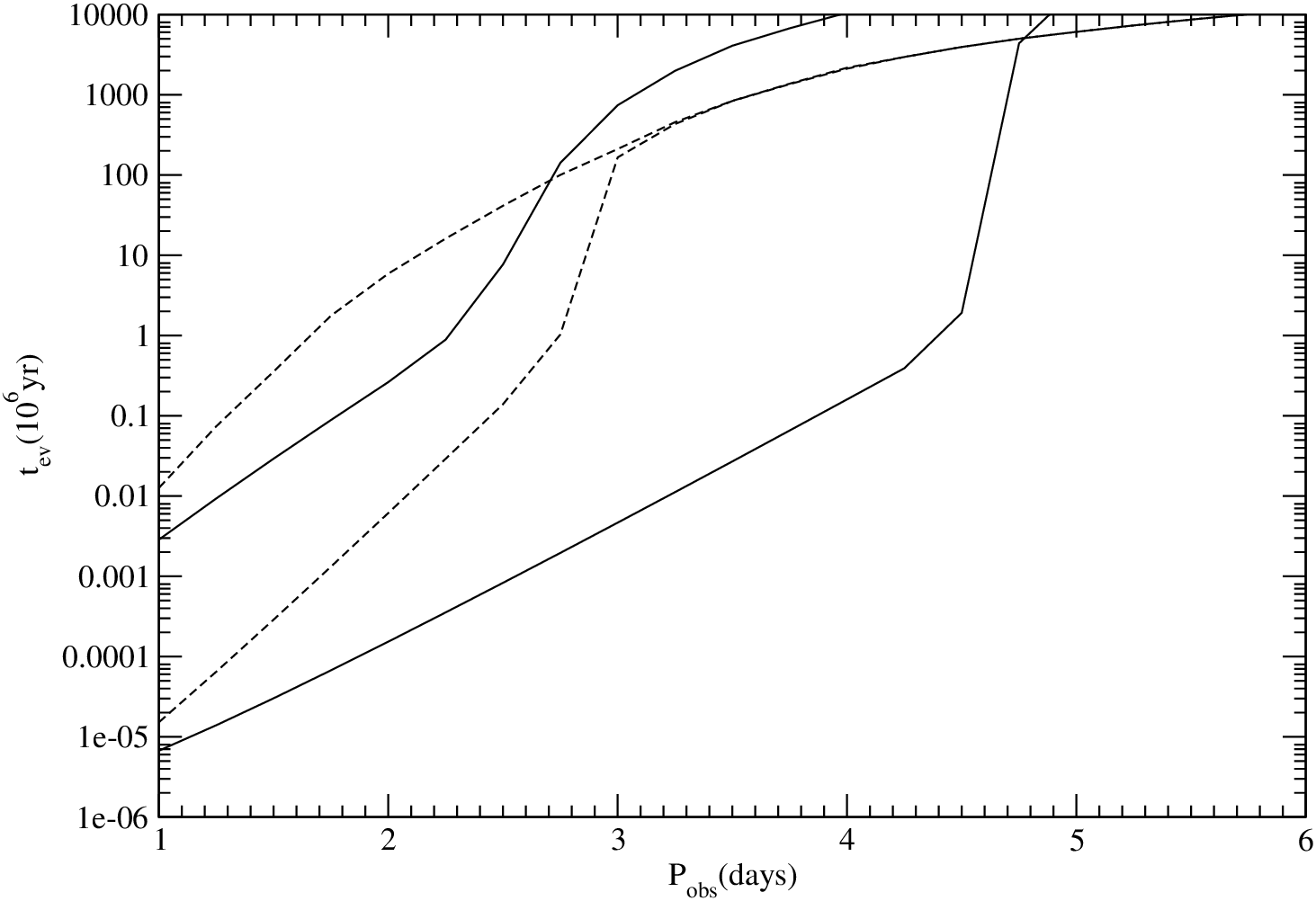}
\caption{The time $t_{ev}$ (see equation (\ref{eqno99})
for definition) is shown as a function of $P_{obs}.$ The solid and dashed curves
correspond to  $m_{pl}=1M_{J},5M_{J}$, respectively. The lower and upper curves of the same type correspond to
($\Omega_{r}=0,\Omega_{crit}$). We choose $a_{in}=10Au$ for all curves.}
\label{Fig14}
\end{figure}

\begin{figure}
\vspace{8cm}\includegraphics{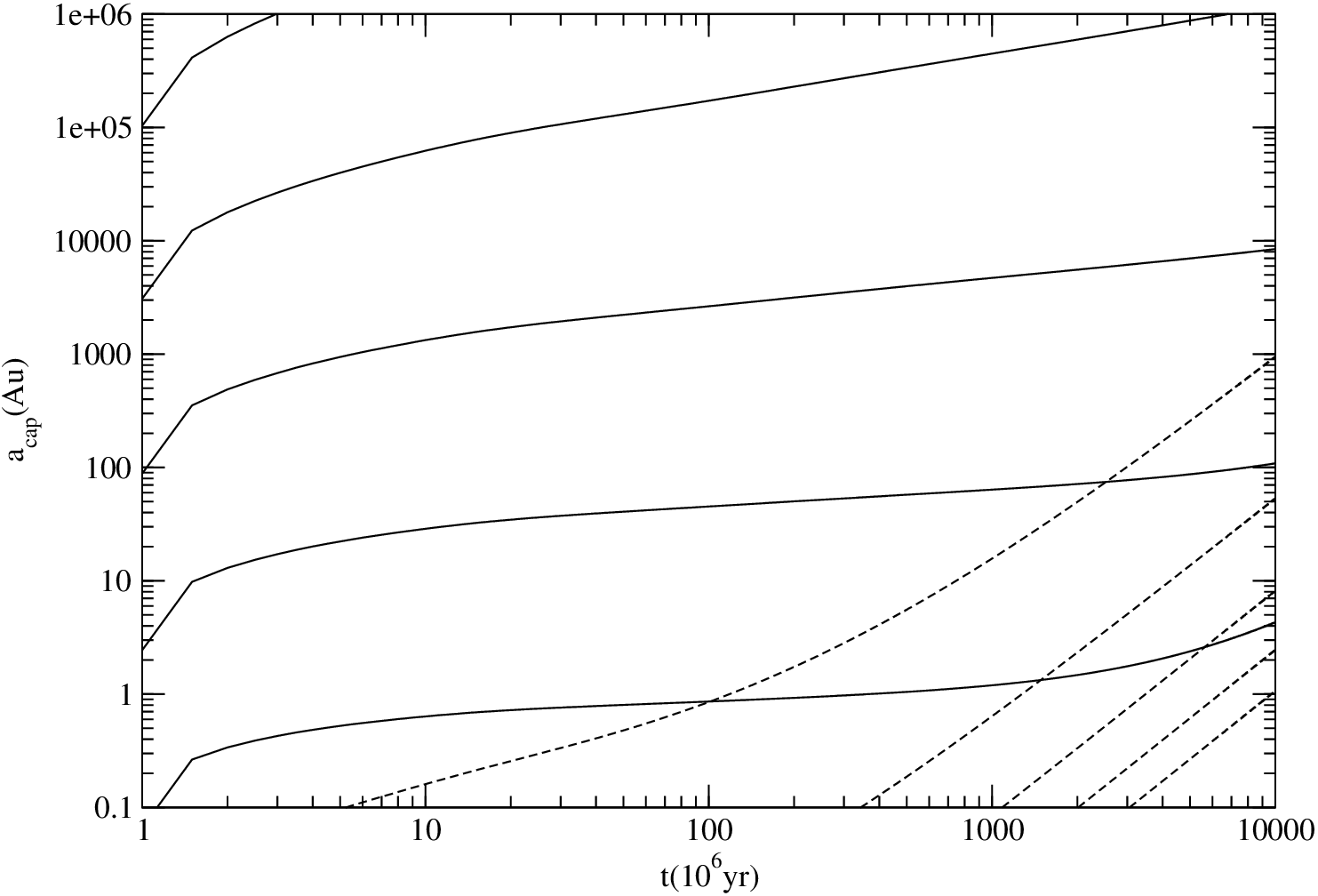}
\caption{The dependence of $a_{cap}$ (in $Au$ ) on time (in units $10^{6}yr)$.
 The solid and dashed curves correspond to
$\Omega_{r}=0,\Omega_{crit}$. The five curves of the same type correspond to different $P_{obs}=3,3.5,4,4.5,5day$.
The lower curves correspond to larger periods.}
\label{Fig15}
\end{figure}

\begin{figure}
\vspace{8cm}\includegraphics{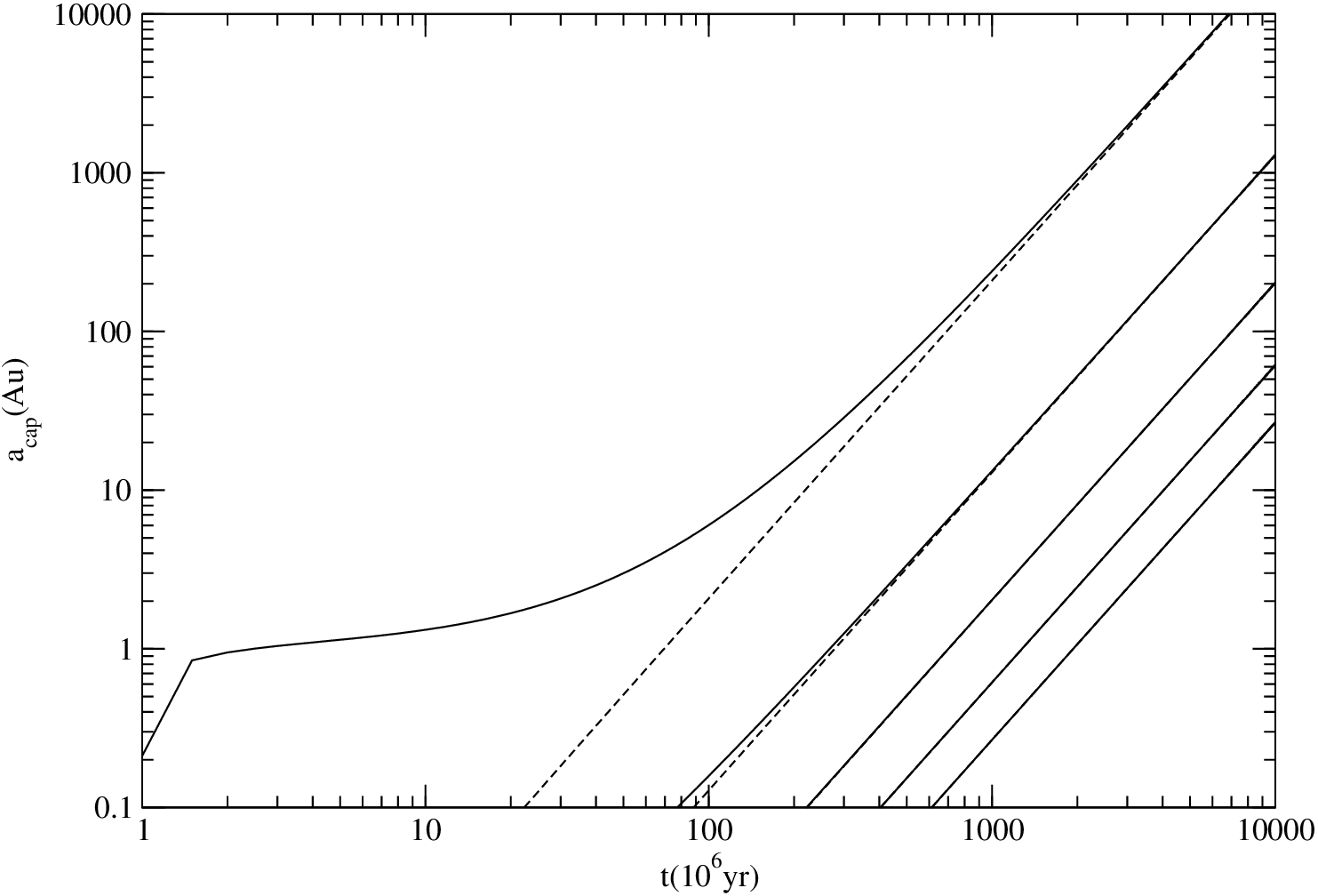}
\caption{Same as figure \ref{Fig15}, but for $m_{pl}=5M_{J}$.
Note that all curves except the curve corresponding to $\Omega_{r}=0$ and
$P_{obs}=3days$ are close to straight lines. Also, curves with the same period $P_{obs}$ are close to each other.
This is due to the fact that the orbital evolution
is mainly determined by dynamic tides in the star for $m_{pl}=5M_{J}$.}
\label{Fig16}
\end{figure}

One can see from figure (14) that for a non-rotating planet and 
$P_{obs}\sim 1day$, $t_{ev}$ is about a few years and is of order of the orbital period.
Such a fast tidal evolution may possibly lead to disruption of the planet (see also Discussion). 

\subsection{Onset of the stochastic instability}

Finally, let us discuss the condition leading to stochastic build-up of the mode energy. From the results of
Section 2.9 it follows
that the stochastic instability sets in when the orbital semi-major axis is sufficiently large: $a \geqsim a_{st}$,
where $a_{st}$ can be expressed in 
term of the parameter $\bar \alpha_{crit}$ (see equation ({\ref{eqno80aa}})) using 
equation (\ref{eqno78})
$$a_{st}=({\alpha_{crit}E_{pl}\over 6\pi\tilde \omega_{00}(\Delta E_{ns}+\Delta E_{ns *})})^{2/5}({M\over m_{pl}})^{3/5}R_{pl}
\approx 30.8(\tilde \omega_{00}\epsilon_{ns})^{-2/5}({M_{J}\over m_{pl}})^{3/5}({R_{pl}\over R_{J}})Au, \eqno 107$$
and we use the  value
$\bar \alpha_{crit}=0.67$ on the right hand side.
\begin{figure}
\vspace{8cm}\includegraphics{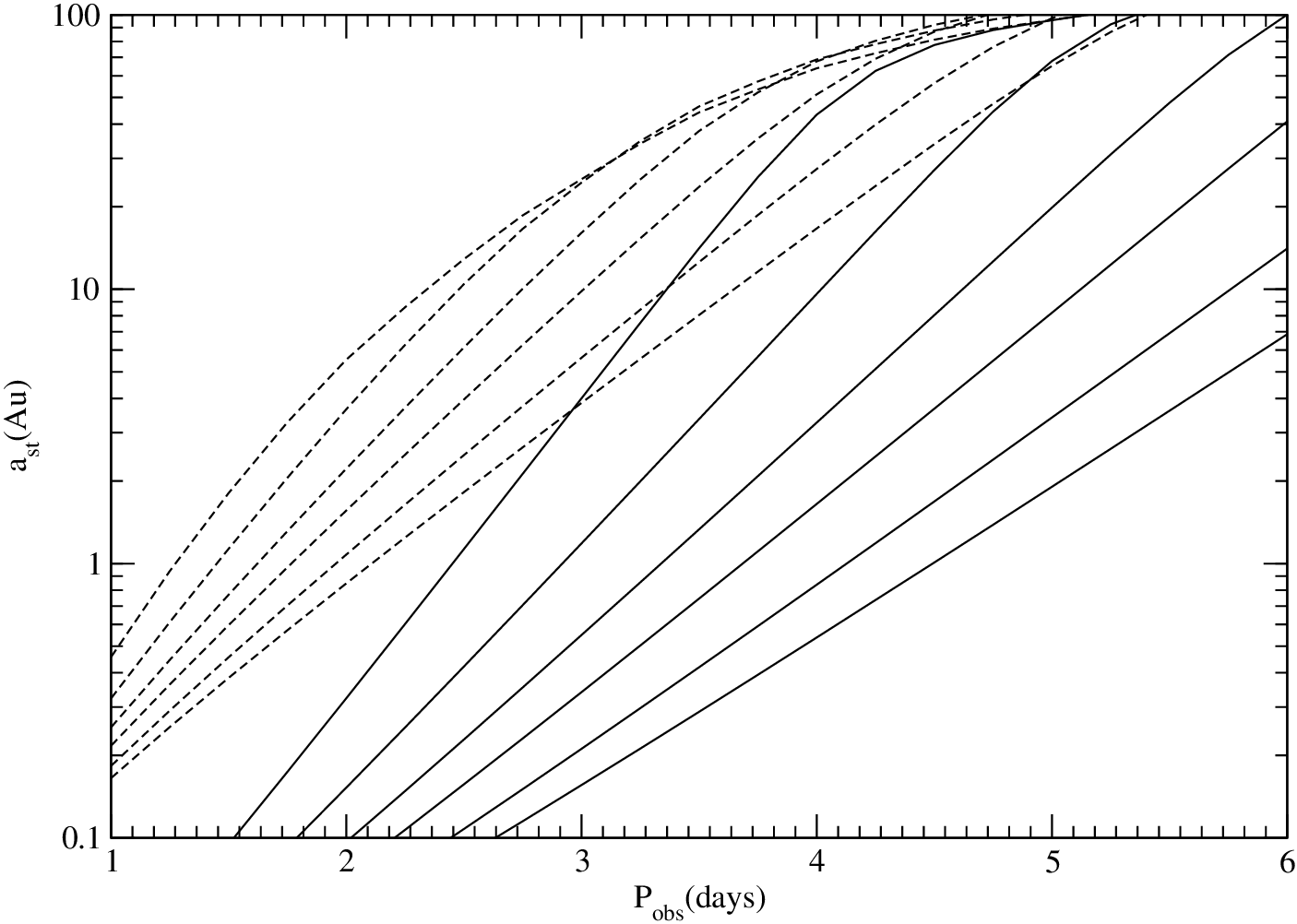}
\caption{The dependence of $a_{st}$ (in $Au$) on $P_{obs}$ (in days). The solid and dashed curves correspond to
$\Omega_{r}=0,\Omega_{crit}$. Six curves of the same type correspond to different $R_{pl}/R_{J}=1,1.2,1.4,1.6,1.8,2$.
The smaller values of $a_{st}$ correspond to larger values of $R_{pl}$.} 
\label{Fig17}
\end{figure}

The result of calculation of $a_{st}$ for a 
$1M_{J}$ planet is shown in figure \ref{Fig17}.
One can see from this that the
'stochastic scale' is of the order of the
scales of interest $\sim 0.1-100Au$, with
larger values corresponding to the planet with $\Omega_{r}=\Omega_{crit}$.

Accordingly, the stochastic build-up of mode energy
may be difficult to achieve for planets rotating at the
critical rate  (see however Section 4).
For non-rotating planets this problem is much less severe.
For example, for the non-rotating planet 
with $R_{pl}\approx 1.6R_{J}$ and $P_{obs}\sim 4days$
we have $a_{st}\approx 2Au$ (figure \ref{Fig17}).
Thus an orbit with 'initial'
value of the semi-major axis $\sim 10Au$  can be evolved by 
stochastic tides in  a  time $t_{ev}\sim 10^{5}yr$
(see figure \ref{Fig14}) which is much smaller than the planet's cooling time $\sim 10^{6}yr$.
Therefore dynamical tides can significantly
change the semi-major axes of non-rotating planets with
 parameters appropriate for observed systems.

It is instructive to compare the analytic condition (107) with the results of numerical
experiments (Mardling 1995 a, Mardling $\&$ Aarseth 2001). Madling (1995)a and Marldling 
$\&$ Aarseth (2001) define the boundary between stochastic and regular regimes of tidal evolution
as a curve $D_{min}^{crit}(e)$ on the plane $(e, D_{min})$, where $e$ is the eccentricity of
the orbit.
The stochastic region is situated below this curve. For a $n=1.5$ polytrope Mardling (1995)a
obtains: $D_{min}^{crit}(e\approx 0.9)\approx 3.75$, 
$D_{min}^{crit}(e\approx 0.95)\approx 4.2$, $D_{min}^{crit}(e\approx 0.99)\approx 4.5$, where 
the minimal separation or pericentre
 distance is expressed in units of stellar radii. Assuming that the mass ratio
is equal to one and using the values $\tilde \omega_{00}\approx 1.45$ and $\tilde Q\approx 0.5$
(Lee $\&$ Ostriker 1986) we can obtain a similar curve $\tilde D_{min}^{crit}(e)$ from
our criterion (107). This curve is qualitatively similar to $D_{min}^{crit}(e)$.
We have $\tilde D_{min}^{crit}(e\approx 0.9)\approx 3.1$, 
$\tilde D_{min}^{crit}(e\approx 0.95)\approx 3.4$,
$\tilde D_{min}^{crit}(e\approx 0.99)\approx 4$. Our curve lies below $D^{crit}_{min}(e)$ on
the plane $(e, D_{min})$  and for a specified value of $e > 0.9$
 gives a value of $D_{min}$ that  is
$10-20$ percent smaller.   A possible reason for this difference is the  neglect of
higher order terms in the tidal potential and 
the contribution of
p-modes in our treatment of the problem which would enhance the tidal
interaction and lead to larger values of $ D_{min}$.

We do not discuss the condition for
stochastic build-up of mode energy for planets with $m_{pl}=5M_{J}$.
As  was shown
above, the semi-major axis of such a planet is mainly evolved
as a result of dynamical tides in the star. These  tides
mainly excite  $f$ and low order $g$ modes.
These modes may interact effectively with high $l$ $g$-modes
due to resonant non-linear mode-mode interaction (Kumar $\&$ Goodman 1996).
Kumar and Goodman estimated the energy in tidally excited
modes needed to be larger than $\sim 10^{36}ergs$
for the resonant parametric instability to occur.
This is of the
order of the energy obtained by the star after a fly-by of a 
$5M_{J}$ planet  with minimal separation distances typical
of our problem.
The parametric instability may not only damp effectively 
the tidally exited modes but also lead to
phase changes in these modes thus 
facilitating the stochastic build-up of  mode energy in the star.
However, this possibility needs further investigation.

\section{Discussion and Conclusions}

In this paper we have 
developed a theory of the tidal interaction between a central 
star and a massive convective planet moving
on a highly eccentric orbit. We showed that 
it is possible to develop a self-consistent perturbative approach
to the theory of disturbances induced in a slowly rotating
planet by tidal interaction by regarding them as resulting
from a sequence of impulsive interactions occurring at periastron (e.g.. Press \& Teukolsky 1977).
We obtained  contributions to the  energy and angular momentum   exchanged  during periastron passage
from both quasi-static
and dynamic tides that were derived 
from  the same set of governing  equations.
Quasi-static tides require the existence of dissipative processes
in order to produce net energy and angular momentum transfers from the orbit.
However, dynamic tides may do this through the excitation
of normal modes.

We  found that if the planet was initially non rotating, dynamic tides
would transfer angular momentum to it until a 'critical' equilibrium prograde
rotation rate, $\Omega_{crit},$ was attained. This is defined by the condition that
dynamic tides acting on a planet rotating with the
angular velocity $\Omega_{crit}$ at periastron result in no
net angular momentum transfer and hence do not change the rotation rate of the planet
\footnote{Obviously, the value of  $\Omega_{crit}$   would change in a more 
detailed treatment of the problem  that takes account  the
contribution of other modes and the influence of quasi-static tides.}.                           }.
We found that the rotation rate of the planet
which minimised the  energy gained  as a result of a  periastron passage
was  also equal to $\Omega_{crit}.$
We showed further that this rotation rate
depends on the periastron distance between planet and star and that it can be 
larger than the similar equilibrium
'pseudo-synchronisation' rotation rate $\Omega_{ps}$ obtained through the action of quasi-static tides.
In fact, for
sufficiently small  periastron distances, our analysis indicates that $\Omega_{crit}$ can attain values 
of the order of the planet characteristic
frequency $\Omega_{*}=\sqrt{{Gm_{pl}\over R_{pl}^{3}}}$
so that a planet rotating with $\Omega_{r}\sim \Omega_{crit}$ may
be near to rotational break up. However, the analysis
assumed slow rotation. Thus this issue requires further investigation and consideration of the 
tidal interaction when the planet rotates rapidly.

We examined the consequences
of multiple periastron passages and energy exchanges, which,  because they occur impulsively at periastron,
could be modelled
with a simple algebraic map.
 Multiple encounters can lead to a  stochastic build-up of the energy contained in normal
modes of oscillation (see Mardling 1995 a,b).
We found a  simple criterion for the  instability of the dynamical system
to such a stochastic build-up.

Our results can be applied not only to
recently discovered planetary systems containing a star and massive planet, but also in  the 
much more general context of the tidal interaction
of a solar type star  with a fully  convective object 
or the tidal interaction between  two fully convective objects (such as e.g. low mass stars, brown dwarfs, etc.).

We focused on the problem of the  tidal evolution of a planet orbiting around a star 
on a highly eccentric orbit. Such a configuration might be produced as a result of 
gravitational interaction in a many planet system (e.g. Weidenschilling \& Mazari 1996; Rasio \& Ford 1996;
Papaloizou \& Terquem 2001; Adams \& Laughlin 2003). 

The minimum periastron distance of the orbit was assumed  to be $\geqsim 4$ stellar radii 
corresponding to  a final  period $P_{obs}$,
after an (assumed) stage of tidal circularization at constant orbital angular momentum,
 exceeding about $3days$. The initial  semi-major
axis $a$ was assumed  to be of the order of some 'typical' planet separation
distance   ranging from $1-100Au.$ 

\noindent We made some
simple estimates of the possible tidal evolution of the orbit. 
We used a realistic model of the planet taking into account
the fact that the radius and luminosity decrease with age as the planet cools.
We first showed  that quasi-static tides cannot 
account for a significant tidal evolution
when the semi-major axes exceed
$ \sim 0.1Au$. 
However, dynamical tides can in principle 
lead to a  large decrease of the semi-major axis
and circularize the  orbit on a time-scale $t_{ev} \leqsim 10^{9}-10^{10}yr$.

Note that for simplicity we did not take into account possible gravitational interaction
of the planet with orbital parameters evolving due to tides  with other planets in the planetary system. 
The possible   characterization of this interaction could be decribed as follows.
If the characteristic
time of gravitational interaction $t_{gr}$ is smaller than $t_{ev}$ the orbital evolution would be
mainly determined by the interaction with other planets. 
The interaction time $t_{gr}$ may  increase significantly if the
planet is scattered into a highly eccentric orbit by another planet
that is itself ejected or  scaterred to large distance. In addition 
$t_{ev}$ sharply decreases with
decrease of orbital semi-major axis. For a sufficiently small semi-major axis
 one would expect that $t_{ev} < t_{gr}$
and the orbital evolution   to be mainly determined by tides.
Such a formulation of the problem is similar to
the well-known loss cone problem in physics of stellar systems around super-massive black holes
and is sufficiently complex to  require analysis by
numerical means.

In a fully convective planet dynamic tides  cause the
excitation of mainly  the 
$l=2$ fundamental mode of pulsation.
This mode has a period
that is typically much smaller than a characteristic time of periastron  passage
corresponding to a typical orbital period $\sim P_{obs}.$ 
The amount of energy and angular momentum gained at periastron 
due to dynamical tides decreases exponentially with $P_{obs}.$
Therefore, the evolution time $t_{ev}$ 
is very sensitive to the value of
$P_{obs}.$ 
Orbits of planets with mass $\sim 1M_{J}$ can be effectively circularized by the tides for
final  orbital periods
$P_{obs}\sim 3-4days$ corresponding to the shortest 
observed periods in extra-solar planetary systems.
However, the contribution
of dynamic tides to the orbital circularization  rate does not 
seem to be effective for larger periods $P_{obs} \geqsim 5days$.
These results appear to be consistent with the observational result
that the orbits of extra-solar planets are nearly circular (eccentricity $< 0.06$)
for periods $P_{obs} \leqsim 5days$.
Dynamic tides raised in the star
are not important for 
the orbital evolution of $1M_{J}$ planets. However,  for  planets with
larger mass $\geqsim 5M_{J}$, the stellar dynamic tides are significant.

There are at least three possible problems with the scenario of tidal circularization we have discussed here.
Firstly, dynamical tides can only
transfer energy from the orbit to the planet effectively
only for sufficiently large orbital semi-major axes $a > a_{st}.$
For orbits with $a < a_{st}$, stochastic instability leading to the energy transfer is 
ineffective so that the orbital
evolution would instead be determined by the decay time of the fundamental mode of oscillation.
This would lead to 
a very large circularization time-scale. 
This difficulty can possibly be overcome in the following way. Suppose
that the planet starts its tidal evolution with $a > a_{st}.$ When $a$
has decreased  to $a\sim a_{st}$, assuming little dissipation, 
a significant energy $\sim {GMm_{pl}\over a_{st}}$  would be contained
in the oscillations of the planet. 
Eventually non linear effects should start to cause
dissipation of this energy  and a significant expansion of the radius which in turn
produces a significant increase  in the
 energy transfer  from the orbit to the planet by tides.
This could produce 
a form of tidal runaway  (e.g.. Gu, Lin $\&$ Bodenheimer 2003)
until possibly  a balance between dissipation of energy
due to nonlinear effects and tidal injection is attained.
In this situation, a condition of effective stochastic instability
may  be fulfilled all the way down to the final quasi-circular orbit.
Self-consistent calculations taking into
account the non-linear effects of tidal heating on the structure of 
the planet are needed in order to  evaluate this possibility
\footnote{Tidal heating in a similar 
context has been discussed by Podsiadlowski (1996) for stellar systems and
more recently by Gu, Lin $\&$ Bodenheimer (2003) for 
 short period planets heated by the action of  quasi-static tides.}.

On the other hand,
if the dynamical tides operate effectively on scales 
$\ll a_{st}$ there is enough energy exchanged to  destroy the planet.
Indeed if the planet semi-major axis is smaller than
$a_{*}={M\over m_{pl}}R_{pl}\approx
7\cdot 10^{12}({M_{J}\over m_{pl}})cm$, the magnitude of the  orbital
energy becomes larger than the internal energy of the planet. 
Therefore, in order to settle onto a tight circular orbit
around the star,   short period planet must dissipate and radiate
away an amount of energy which  is much  larger than  its 
internal energy. 
This issue needs to be discussed in the context of 
a non-linear analysis of the orbital evolution.

Finally,  in our approximations, the influence of dynamical tides decreases exponentially with
$P_{obs}$ and the tidal evolution as well as initial conditions of the problem
are likely to be stochastic. 
Under such conditions there may exist other
planets in the same system which have failed to circularize their orbits and 
have very high orbital eccentricities at the present time
(the difference $1-e\sim 10^{-3}-10^{-4}$).
This difficulty  may appear less significant when one considers the  possible contribution
of other non-standard long period modes of oscillation.
The energy  transferred to these modes could lead to tidal evolution
at larger separation distances (corresponding to $P_{obs} \geqsim 5$).

There are at least two possible candidates for such
modes. Firstly, there is a low-frequency mode associated with the  first order phase transition between molecular
and metallic Hydrogen (Vorontsov 1984, 
Vorontsov, Gudkova and Zharkov 1989). This mode has a period approximately twice
as long as the period of the fundamental mode. However this mode is probably not important for our problem.
As we have mentioned before the phase transition is absent in hot young planets. Also, this mode disappears when
the time scale of the transition of one phase to another is smaller than the period of the mode  (Vorontsov 1984).
Another very interesting possibility is to consider the inertial modes associated with rotation of the planet
(e.g. Papaloizou $\&$ Pringle 1978). The modes have very low frequency and therefore their excitation
does not decrease exponentially with distance.
The density perturbation associated with these modes is small being of order $ \Omega_{r}^{2}.$
However, the contribution of these modes to the energy gained from the orbit
may be significant for large values of $P_{obs}$
where the contribution of the fundamental mode is exponentially small. We will to consider the excitation
of these modes by dynamical tides in our future work.

\section*{Acknowledgements}
We are grateful to T. V. Gudkova, R. P. Nelson, F. Vivaldi and V. N. Zharkov for useful remarks. It is our sincere pleasure
to thank S. V. Vorontsov for many very helpful discussions.

\appendix
\section[]{Derivation of the explicit expression for the damping rate}

In order to obtain equation (\ref{eqno17}) from equation (\ref{eqno14}),
we need to evaluate the following expression
$$a={1\over 2}\int d\Omega \sigma^{*\beta}_{\alpha}\sigma^{\alpha}_{\beta}, \eqno A1$$
where $\sigma^{*\beta}_{\alpha}$ is related to the displacement
through equation (\ref{vstr}) and  integration is performed over the solid angle $\Omega$.
We assume  an  isentropic condition  leading to a circulation  free
displacement such that  $\nabla \times \bvec {\xi} =0$ is fulfilled.
We may then  introduce a  displacement potential $\Phi$ such  that 
${\bvec{ \xi}}=\nabla \Phi$, and rewrite (A1) in terms of it so obtaining
$$a=2\int d\Omega (\Phi^{*}_{,\alpha \beta}\Phi_{,\alpha \beta }-{1\over 3}\Delta \Phi^{*} \Delta \Phi). \eqno A2$$
We use the identity
$$\Phi^{*}_{,\alpha \beta}\Phi_{,\alpha \beta}={1\over 2}(\Delta (\Phi^{*}_{,\alpha} \Phi_{,\alpha})
-(\Phi^{*}_{,\alpha}\Delta \Phi_{,\alpha}+\Phi_{,\alpha}\Delta \Phi^{*}_{,\alpha})),$$
to bring (A2) in the form
$$a=\int d\Omega  (\Delta (\Phi^{*}_{,\alpha} \Phi_{,\alpha})
-(\Phi^{*}_{,\alpha}\Delta \Phi_{,\alpha}+\Phi_{,\alpha}\Delta \Phi^{*}_{,\alpha}) 
-{2\over 3}\Delta \Phi^{*} \Delta \Phi)$$
$$=\int d\Omega (\Delta (\bvec { \xi}\cdot \bvec { \xi}^{*}) 
-\bvec { \xi}^{*}\cdot \nabla (\nabla \cdot \bvec {\xi})
-\bvec {\xi}\cdot \nabla (\nabla \cdot \bvec {\xi}^{*})
-{2\over 3}(\nabla \cdot \bvec { \xi})(\nabla \cdot \bvec {\xi}^{*})). \eqno A3$$
Now we substitute equation (\ref{eqno11a}) into (A3), use the relation (\ref{eqno12}) and
the known properties of the vector functions
to obtain
$$a=2\lbrace {2\over 3} (\xi_{R}^{'}-{\xi_{R}\over r}+{L^{2}\over 2}{\xi_{S}\over r})^{2}+
{1\over r^{2}}(2(\xi_{R}-{L^{2}\over 2}\xi_{S})^{2}+
(L^{2}-2)\xi_{R}^{2}+L^{2}{(\xi_{R}-\xi_{S})}^{2})\rbrace, \eqno A4$$
where $L^{2}=l(l+1)$. Note that (A4) cannot be used for $l=0$.
Multiplying equation (A4) on ($r^{2} \rho \nu$), setting $l=2$ ($L^{2}=6$)
and integrating the result over $r$ we get equation (\ref{eqno17}). 

\bsp

\label{lastpage}

\end{document}